\newcommand*{\ATLASLATEXPATH}{latex/}
\documentclass[cernpreprint,texlive=2011,txfonts,UKenglish]{latex/atlasdoc}
\pdfoutput=1
\usepackage{latex/atlaspackage}
\usepackage{latex/atlasphysics}

\usepackage{\ATLASLATEXPATH atlaspackage}
\usepackage{\ATLASLATEXPATH atlasbiblatex}

\usepackage{\ATLASLATEXPATH atlascontribute}

\usepackage{\ATLASLATEXPATH atlasphysics}

\usepackage{multirow}
\usepackage{siunitx}

\graphicspath{{logos/}{figures/}}

\usepackage{Paper01-defs}

\AtlasTitle{Search for Minimal Supersymmetric Standard Model Higgs bosons 
  $H/A$ and for a $Z^{\prime}$ boson in the $\tau \tau$ final state produced in $pp$ collisions at $\sqrt{s}=13$~TeV
  with the ATLAS Detector}

\author{The ATLAS Collaboration}

\date{\today}

\AtlasRefCode{HIGG-2015-08}

\PreprintIdNumber{CERN-EP-2016-164}

\AtlasJournalRef{Eur. Phys. J. C76 (2016) 585}
\AtlasDOI{10.1140/epjc/s10052-016-4400-6}

\AtlasJournal{Eur.\ Phys.\ J.\ C.}

\AtlasAbstract{%
A search for neutral Higgs bosons of the Minimal Supersymmetric Standard Model (MSSM) and for a 
heavy neutral $Z^{\prime}$ boson is performed using a data sample corresponding to an integrated 
luminosity of 3.2~fb$^{-1}$ from proton--proton collisions at $\sqrt{s} = 13$~\TeV\ recorded by the 
ATLAS detector at the LHC.  The heavy resonance is assumed to decay to a $\tau^+ \tau^-$ pair with 
at least one $\tau$ lepton decaying to final states with hadrons and a neutrino. The search is performed in the mass range of 
0.2--1.2~\TeV\ for the MSSM neutral Higgs bosons and 0.5--2.5~\TeV\ for the heavy neutral 
$Z^{\prime}$ boson.  The data are in good agreement with the background predicted by the Standard 
Model. The results are interpreted in  MSSM and $Z^{\prime}$ benchmark scenarios.
The most stringent constraints on the MSSM $m_A$--$\tan\beta$ space exclude at 
95\% confidence level (CL) $\tan\beta > 7.6$ for $m_A = 200$~\GeV\ in the $m_{h}^{\text{mod+}}$ MSSM scenario. 
For the Sequential Standard Model, a $Z^{\prime}_\mathrm{SSM}$ mass up to 1.90~\TeV\ is excluded 
at 95\% CL and masses up to 1.82--2.17~\TeV\ are excluded for a $Z^{\prime}_{\mathrm{SFM}}$ of the Strong 
Flavour Model. 
}

\hypersetup{pdftitle={ATLAS document},pdfauthor={The ATLAS Collaboration}}

\begin{document}
\maketitle
\tableofcontents
\section{Introduction}
\label{sec:intro}

The discovery of a scalar particle at the LHC \cite{ATLASHiggsJuly2012, CMSHiggsJuly2012} has
provided important insight into the mechanism of electroweak symmetry breaking.  Experimental
studies of the new particle
\cite{Aad:2015mxa,Aad:2015gba,Khachatryan:2014jba,Khachatryan:2014kca,ATLAS-CONF-2015-044}
demonstrate consistency with the Standard Model (SM) Higgs boson \cite{ENGLERT,HIGGS,HIGGS2,HIGGS3,
  Guralnik:1964eu,Kibble:1967sv}. However, it remains possible that the discovered particle is part
of an extended scalar sector, a scenario that is favoured by a number of theoretical arguments
\cite{Djouadi:2005gj,Branco:2011iw}.

The Minimal Supersymmetric Standard Model (MSSM)~\cite{Fayet:1976et,Fayet:1977yc,Farrar:1978xj,
  Fayet:1979sa,Dimopoulos:1981zb} is the simplest extension of the SM that includes
supersymmetry. The MSSM requires two Higgs doublets of opposite hypercharge.  Assuming that CP symmetry is
conserved, this results in one CP-odd ($A$) and two CP-even ($h$, $H$)  neutral Higgs bosons and two
charged Higgs bosons ($H^{\pm}$).  At tree level, the properties of the Higgs sector in the MSSM depend on only
two non-SM parameters, which can be chosen to be the mass of the CP-odd Higgs boson, $m_A$, and the
ratio of the vacuum expectation values of the two doublets, $\tan\beta$. Beyond tree level,
additional parameters affect the Higgs sector, the choice of which defines
various MSSM benchmark scenarios. In some scenarios, such as  $m_{h}^{\text{mod}+}$~\cite{MSSMBenchmarks},
the top-squark mixing parameter is chosen such that
the mass of the lightest CP-even Higgs boson, $m_h$, is close to the measured mass of the Higgs
boson that was discovered at the LHC.
A different approach is employed in the hMSSM scenario~\cite{Djouadi:2013uqa,Bagnaschi:2015hka} in which
the value of $m_h$ can be used, with certain assumptions, to predict the remaining masses and couplings of the MSSM Higgs bosons without explicit reference to the soft supersymmetry-breaking parameters.
The couplings of the MSSM heavy Higgs bosons to down-type fermions are enhanced with respect to the
SM for large $\tan\beta$ values, resulting in increased branching fractions to $\tau$ leptons and
$b$-quarks,\footnote{Throughout this paper the inclusion of charge-conjugate decay modes is implied.}
as well as a higher cross section for Higgs boson production in association with $b$-quarks.
This has motivated a variety of searches for a scalar boson in $\tau\tau$ and $bb$ final states at
LEP~\cite{LEPLimits}, the Tevatron~\cite{TevatronLimits1,TevatronLimits2,TevatronLimits3} and the
LHC~\cite{Aad:2014vgg,ATLASLimit,Khachatryan:2014wca,Khachatryan:2015tra,LHCbHtautau}.

Heavy \Zprime{} gauge bosons appear in several
models~\cite{Hewett:1989,Cvetic:1995zs,Leike:1998wr,Rizzo:2006nw,Diener:2010sy} and are a common
extension of the SM~\cite{Langacker:2008}.  Such \Zprime{} bosons can appear in theories extending
the electroweak gauge group, where lepton universality is typically conserved.  A frequently used
benchmark is the Sequential Standard Model (SSM)~\cite{SSM}, which contains a single additional \Zprime{} boson
with the same couplings as the SM $Z$ boson.  Some models offering an explanation for the high mass
of the top-quark, predict instead that such bosons couple preferentially to third-generation
fermions~\cite{ZprimeToThirdGen,sfm,G221,TopFlavor}.  A model predicting additional weak gauge
bosons \Zprime{} and \Wprime{} coupling preferentially to third-generation fermions is the Strong
Flavour Model (SFM)~\cite{sfm,TopFlavor}.

\begin{figure}[t!]
  \centering
  \subfloat[]{\includegraphics[width=0.25\textwidth]{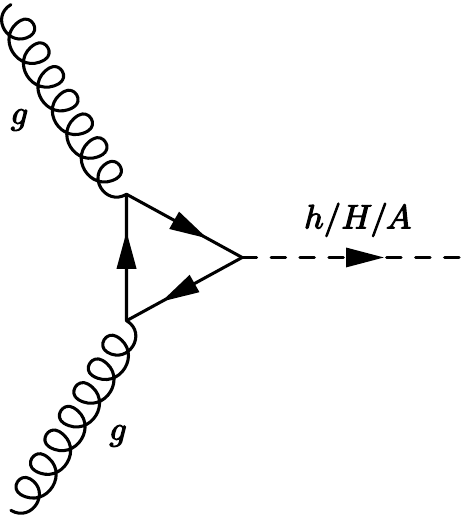} }
  \hspace{1.5cm}  
  \subfloat[]{\includegraphics[width=0.25\textwidth]{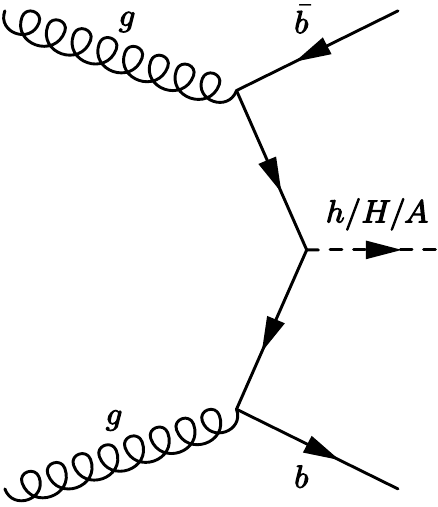} }\\
  \subfloat[]{\includegraphics[width=0.25\textwidth]{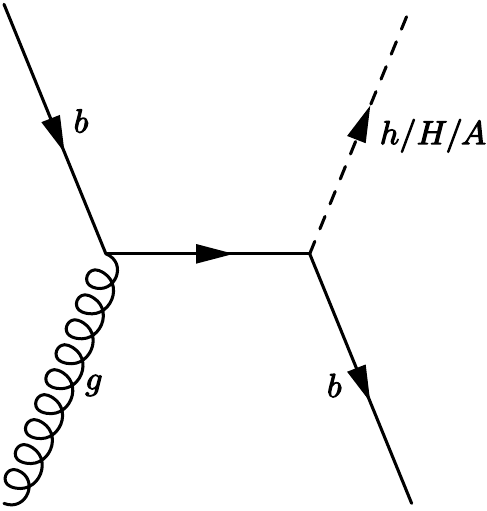} }
  \hspace{1.5cm}
  \subfloat[]{\includegraphics[width=0.25\textwidth]{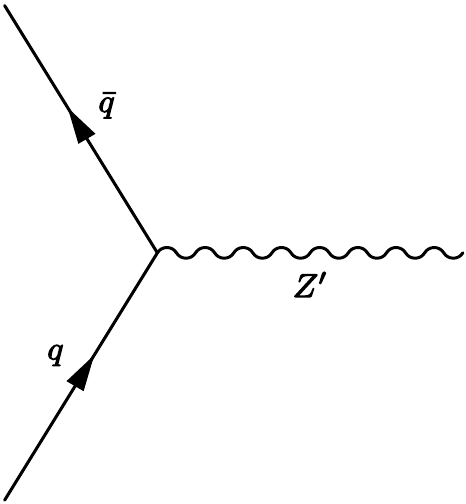} }
  \caption{Lowest-order Feynman diagrams for (a) gluon--gluon fusion  and $b$-associated production in the
    (b) four-flavour  and (c) five-flavour  schemes of a neutral MSSM Higgs boson.   Feynman diagram
    for Drell--Yan production of a \Zprime{} boson at lowest order  (d).}
\label{fig:feynman}
\end{figure}

Direct searches for high-mass  resonances decaying to $\tau\tau$ have been performed by the
ATLAS~\cite{ATLASZprime5fb} and CMS~\cite{CMSZprime5fb} collaborations using 5~\ifb{} of integrated
luminosity at $\sqrt{s}=7~\TeV$.
ATLAS~\cite{ATLASZprime20fb} updated the search with 20~\ifb{} of integrated
luminosity at $\sqrt{s}=8~\TeV$.
Indirect limits on \Zprime{} bosons with non-universal flavour couplings have been set based on
measurements from LEP~\cite{ZprimeMassLimitsFromLEP}.

This paper presents the results of a search for neutral MSSM Higgs bosons as well as high-mass
\Zprime{} resonances in the $\tau\tau$ decay mode using \currentlumirange{} of proton--proton
collision data collected with the ATLAS detector \cite{ATLASDetector} in 2015 at a centre-of-mass
energy of 13~\TeV.  The search is performed for the $\taulep\tauhad$ and $\tauhad\tauhad$ decay modes,
where $\taulep$ represents the decay of a $\tau$ lepton to an electron or a muon and neutrinos and 
$\tauhad$ represents the decay to one or more hadrons and a neutrino. 
The search considers narrow resonances in the mass range of 0.2--1.2~\TeV{} and $\tan\beta$ range of 1--60 for the MSSM Higgs bosons. For the \Zprime{} boson search, the mass range of 0.5--2.5~\TeV{} is considered.
Higgs boson production through gluon--gluon fusion and in association with $b$-quarks is considered
(Figures~\ref{fig:feynman}(a)--\ref{fig:feynman}(c)), with the latter mode dominating for high $\tan\beta$ values.  Hence
both the $\taulep\tauhad$ and $\tauhad\tauhad$ channels are split into $b$-tag and $b$-veto
categories, based on the presence or absence of jets originating from $b$-quarks in the final state.  Since a
\Zprime{} boson is expected to be predominantly produced via a Drell--Yan process
(Figure~\ref{fig:feynman}(d)), there is little gain in splitting the data into $b$-tag and $b$-veto
categories.  Hence, the \Zprime{} analysis uses an inclusive selection instead.

\section{Data sample and Monte Carlo simulation}
\label{sec:samples}
The ATLAS detector~\cite{ATLASDetector} at the LHC consists of an inner tracking detector with a
coverage in pseudorapidity\footnote{ATLAS uses a right-handed coordinate system
  with its origin at the nominal interaction point (IP) in the centre of the detector and the
  $z$-axis along the beam pipe. The $x$-axis points from the IP to the centre of the LHC ring, and
  the $y$-axis points upwards. Cylindrical coordinates $(r,\phi)$ are used in the transverse plane,
  $\phi$ being the azimuthal angle around the beam pipe. The pseudorapidity is defined in terms of
  the polar angle $\theta$ as $\eta=-\ln\tan(\theta/2)$. Angular distance is measured in units of
  $\Delta R \equiv \sqrt{(\Delta\eta)^2 + (\Delta\phi)^2} $.}  up to $|\eta|~=~2.5$ surrounded by a
thin superconducting solenoid providing a 2~T axial magnetic field, electromagnetic and hadronic
calorimeters extending up to $|\eta|~=~4.9$ and a muon spectrometer covering $|\eta|~<~2.7$.
A new innermost layer was added to the pixel tracking detector after the end
Run-1 at a radial distance of 3.3~cm from the beam line~\cite{ATLAS-TDR-19}.
The ATLAS trigger system consists of a hardware-based first level trigger, followed by a software-based high-level trigger (HLT).
The integrated luminosity used in this search, considering the data-taking periods of 2015 in
which all relevant detector subsystems were operational, is \currentlumirange{}.
The luminosity measurement and its uncertainty are derived following a methodology similar to that detailed 
in Ref.~\cite{lumi}, from a calibration of the luminosity scale using $x$--$y$ beam-separation scans performed in August 2015.

Simulated events with a heavy neutral MSSM Higgs boson produced via gluon--gluon fusion and in association with $b$-quarks
are generated with the {\tt POWHEG-BOX v2}~\cite{Nason:2004rx,Frixione:2007vw,Powheg3} and 
{\tt MADGRAPH5\_aMC@NLO~2.1.2}~\cite{Alwall:2014hca,Wiesemann:2014ioa} programs, respectively.  The
CT10~\cite{Lai:2010vv} and CT10nlo\_nf4~\cite{Dulat:2015mca} sets of parton distribution functions (PDFs) are used,
respectively.  {\tt PYTHIA~8.210}~\cite{PYTHIA82} with the AZNLO~\cite{AZNLO:2014} (A14~\cite{ATL-PHYS-PUB-2014-021}) set of tuned
parameters, or ``tune'', is used for the parton shower, underlying event and hadronization in the
gluon--gluon fusion ($b$-associated) production.  The production cross sections for the
various MSSM scenarios are calculated using {\tt SusHi}~\cite{Harlander:2012pb} for gluon
fusion production~\cite{Harlander:2002wh,Spira:1995rr,Anastasiou:2002yz,Ravindran:2003um,Harlander:2002vv,Anastasiou:2002wq,Aglietti:2004nj,Bonciani:2010ms,Harlander:2004tp,Harlander:2005rq,Degrassi:2010eu,Degrassi:2011vq,Degrassi:2012vt} 
and $b$-associated production in the five-flavour scheme~\cite{Harlander:2003ai};
$b$-associated production in the four-flavour scheme is calculated according to
Refs.~\cite{Dittmaier:2003ej, Dawson:2004a}. The final $b$-associated production cross section
is obtained by using the method in Ref.~\cite{SantanderMatching} to match the four-flavour and
five-flavour scheme cross sections.  The masses and the couplings of the Higgs bosons are computed with
{\tt FeynHiggs}~\cite{Heinemeyer:1998yj, Heinemeyer:1998np, Frank:2006yh, Degrassi:2002fi,
  Hahn:2013ria}, whereas the branching fraction calculation follows the procedure described in
Ref.~\cite{Heinemeyer:2013tqa}.  In the case of the hMSSM scenario, the procedure described in
Ref.~\cite{Bagnaschi:2015hka} is followed for the production cross sections and {\tt HDECAY}
\cite{Djouadi:1997yw} is used for the branching fraction calculation.

The $\Zprime{}$ signals are simulated by reweighting a leading-order (LO) \DY{} sample using the
\texttt{TauSpinner} algorithm~\cite{Czyczula:2012ny,Kaczmarska:2014eoa,TauSpinnerBSM} to account for
spin effects in the $\tau$ decays.  The \DY{} sample, enriched with high invariant mass events, is
generated with \texttt{ PYTHIA~8.165}~\cite{Pythia8} using the NNPDF2.3LO PDF set~\cite{Ball:2012cx}
and the A14 tune for the underlying event.
Interference between the \Zprime{} signals and the SM \DYonly{} production is not included.

The simulated backgrounds consist of the production of $Z$+jets, $W$+jets, $t\bar{t}$ pairs, single
top quarks and electroweak dibosons ($WW/WZ/ZZ$).  These are modelled with several event generators
as described below, while contributions from multi-jet production are estimated with data as described in
Section~\ref{sec:background}.

Simulated samples of $Z$+jets events for the $\taulep\tauhad$ and $\tauhad\tauhad$ channels and
$W$+jets events for the $\taulep\tauhad$ channel are produced using {\tt POWHEG-BOX v2} 
interfaced to {\tt PYTHIA 8.186} with the AZNLO tune. 
In this sample, {\tt PHOTOS++ v3.52}~\cite{photos,Davidson:2010ew} 
is used for final-state QED radiation.
A dedicated $W$+jets sample binned in $\pt^{W}$, produced using the {\tt SHERPA 2.1.1} generator~\cite{Gleisberg:2008ta},
is used in the $\tauhad\tauhad$ channel in order to enhance the number of events with high
invariant mass. For this sample, matrix elements are calculated for up to two partons 
at next-to-leading order (NLO) and four partons at LO, merged with the {\tt SHERPA} 
parton shower model using the {\tt  ME+PS@NLO} prescription~\cite{Hoeche:2012yf}.
Spin correlation effects between the $W$ boson and its decay products are simulated with
the {\tt TauSpinner} program. All $W$/$Z$+jets samples use the CT10 PDF set and are normalized to the
next-to-next-to-leading-order (NNLO) cross sections calculated using {\tt FEWZ}
\cite{Anastasiou:2003ds,Melnikov:2006kv,Gavin:2010az}.

The {\tt POWHEG-BOX v2} program with the CT10 PDF set is used for the generation of $t\bar{t}$ pairs
and single top quarks in the $Wt$- and $s$-channels.  Samples of $t$-channel single-top-quark events
are produced with the {\tt POWHEG-BOX v1} generator employing the four-flavour scheme for the NLO
matrix element calculations together with the fixed four-flavour scheme PDF set CT10f4; the
top-quark decay is simulated with {\tt MadSpin}~\cite{Artoisenet:2012st}.  For all samples of
top-quark production, the spin correlations are preserved and the parton shower, fragmentation and
underlying event are simulated using {\tt PYTHIA 6.428}~\cite{Pythia} with the CTQ6L1 PDF set and
the corresponding Perugia 2012 tune~\cite{Skands:2010ak}. Final-state QED radiation is simulated
using {\tt PHOTOS++ v3.52}.
The top-quark mass is set to 172.5~\GeV.
The $t\bar{t}$ production sample is normalized to the NNLO cross section, including soft-gluon
resummation to next-to-next-to-leading-logarithm accuracy (Ref. \cite{Czakon:2011xx} and references
therein). The normalization of the single-top-quark event samples uses an approximate NNLO calculation from
Refs.~\cite{Kidonakis:2011wy,Kidonakis:2010tc,Kidonakis:2010ux}.

Finally, diboson processes are simulated using the {\tt SHERPA 2.1.1} program with the CT10
PDF. They are calculated for up to one additional parton at NLO, depending on the process, and up to
three additional partons at LO. The diboson samples use the  NLO cross sections {\tt SHERPA} calculates.

The simulation of $b$- and $c$-hadron decays for all samples, excluding those generated with 
{\tt SHERPA}, uses {\tt EvtGen v1.2.0}~\cite{evtgen}. 
All simulated samples include the effect of multiple proton-proton interactions in the same
and neighbouring bunch crossings (``pile-up'') by overlaying simulated minimum-bias events
on each generated signal or background event. These minimum-bias events are generated with 
{\tt PYTHIA 8.186}~\cite{Pythia,Pythia8}, 
using the  A2 tune~\cite{ATLAS:2012uec} and the MSTW2008LO PDF~\cite{Martin:2009iq}.
Each sample is simulated using the full 
{\tt GEANT4}~\cite{Geant4,ATLASSIM} simulation of the ATLAS detector, with the exception of the
$b$-associated MSSM Higgs boson signal, for which the 
{\tt ATLFAST-II}~\cite{ATLASSIM,ATLAS:1300517} fast simulation framework is used.
Finally, the Monte Carlo (MC) samples are processed through the same
reconstruction software as for the data.

\section{Object reconstruction and identification}
\label{sec:reconstruction}

The primary vertex of each event is chosen as the proton–proton vertex
candidate with the highest sum of the squared transverse momenta of all associated tracks.
Electron candidates are reconstructed from energy deposits in the electromagnetic calorimeter
associated with a charged-particle track measured in the inner detector.  The final electron
candidates are required to pass the ``loose'' likelihood-based identification selection
\cite{Aad:2014nim,ATLAS-CONF-2016-024}, to have a transverse energy $E_{\text{T}} > 15$~$\GeV$ and to be in the fiducial
volume of the inner detector, $|\eta|<2.47$.  The transition region between the barrel and end-cap
calorimeters ($1.37<|\eta|<1.52$) is excluded. 

Muon candidates are reconstructed from track segments in the muon spectrometer, matched with tracks
found in the inner detector within $|\eta|<2.5$.  The tracks of the final muon candidates are refit
using the complete track information from both detector systems and are required to have a
transverse momentum $\pt>15$~$\GeV$ and to pass the ``loose'' muon identification
requirements~\cite{ATLASRun2Muon}.

Both the electrons and muons are required to pass a $\pt$-dependent isolation
selection, which utilizes both calorimetric and tracking information, with
an efficiency of 90\% (99\%) for transverse momentum of $\pt=25~(60)$~$\GeV$.
The isolation provides an efficiency that grows as a function of
lepton $\pt$, since the background from jets faking leptons becomes less important as the lepton
$\pt$ increases. The contributions from pile-up and the underlying event activity are corrected on an
event-by-event basis using the ambient energy density technique~\cite{IsolationPileUp1}.

Jets are reconstructed from topological clusters \cite{Aad:2016upy} in the calorimeter using the
anti-$k_t$ algorithm~\cite{AntiKT}, with a radius parameter value $R =0.4$.  To reduce the effect of
pile-up, a jet vertex tagger algorithm is used for jets with $\pt < 50$~$\GeV$ and $|\eta| < 2.4$.
It employs a multivariate technique based on jet energy, vertexing and tracking variables to
determine the likelihood that the jet originates from pile-up~\cite{ATLASJVT}.  In order to identify
jets containing $b$-hadrons ($b$-jets), a multivariate algorithm is used
~\cite{ATL-PHYS-PUB-2015-022,ATL-PHYS-PUB-2015-039}. A working point that corresponds to an average
efficiency of 70\% for $b$-jets in $t\bar{t}$ simulated events is chosen. The misidentification
rates for $c$-jets, $\tau$-jets and jets initiated by light quarks or gluons 
for the same working point and in the same sample of simulated $t\bar{t}$ events 
are approximately 10\%, 4\% and 0.2\% respectively.

Hadronic decays of $\tau$ leptons are predominantly characterized by the presence of one or three
charged particles, accompanied by a neutrino and possibly neutral pions. The reconstruction of the
visible decay products, hereafter referred to as $\tauhadvis$, starts with jets with $\pt>10$~$\GeV$.
The $\tauhadvis$ candidate must have energy deposits in the calorimeters in the range $|\eta| < 2.5$, with the transition
region between the barrel and end-cap calorimeters excluded. Additionally, they must have $\pt > 20~\GeV$, one or three associated tracks and an electric charge of $\pm 1$. A multivariate boosted
decision tree (BDT) identification, based on calorimetric shower shape and track multiplicity of the
$\tauhadvis$ candidates, is used to reject backgrounds from jets.  In
this analysis, two $\tauhadvis$ identification criteria are used: ``loose'' and ``medium'' with
efficiencies measured in $Z\to \tau \tau$ decays of about 60\% (50\%) and
55\% (40\%) for one-track (three-track) $\tauhadvis$ candidates, respectively~\cite{ATLAS:tauperfrun2}.
An additional dedicated
likelihood-based veto is used to reduce the number of electrons misidentified as $\tauhadvis$.

Signals in the detector can be used in more than one reconstructed object.
Objects that have a geometric overlap are removed according to the following priorities: 

\begin{itemize}
\item Jets within a $\Delta R = 0.2$ cone around 
  a selected $\tauhadvis$ are excluded.
\item Jets within a $\Delta R = 0.4$ cone around an electron or muon are excluded. 
\item Any $\tauhadvis$ within a $\Delta R = 0.2$ cone around an electron or muon is excluded.
\item Electrons within a $\Delta R$ = 0.2 cone around a muon are excluded.
\end{itemize}

The missing transverse momentum ($\met$) is calculated as the modulus of the 
negative vectorial sum of the $\vec{\pt}$ of all fully reconstructed and 
calibrated jets and leptons~\cite{ATL-PHYS-PUB-2015-027}. This procedure includes a ``soft
term'', which is calculated based on the inner-detector tracks originating
from the primary vertex that are not associated to
reconstructed objects.

\section{Search channels}
\label{sec:selection}

\subsection{$\taulep\tauhad$ channel}
\label{sec:lephad_selection}

Events in the $\taulep\tauhad$ channel are recorded using single-muon triggers and a logical-OR combination of
single-electron triggers. Single-electron triggers with $\pt$ thresholds of $24 \GeV$, $60 \GeV$ and
$120 \GeV$ are used for the $\tau_e\tauhad$ channel. For the $\tau_\mu\tauhad$ channel, a single-muon trigger
with a $\pt$ threshold of $50~\GeV$ is used if the muon $\pt$ is larger than $55~\GeV$ and  
a single-muon trigger with a $\pt$ threshold of $20~\GeV$ is used otherwise.  The triggers impose electron
and muon quality requirements which are tighter for the triggers with lower $\pt$ thresholds.

Events must have at least one identified $\tauhadvis$ candidate and either one electron or one muon
candidate which is geometrically matched to the HLT object that triggered the event.  Events with
more than one electron or muon fulfilling  the criteria described in Section~\ref{sec:reconstruction}
are rejected in order to reduce the backgrounds from $Z/\gamma^{*}\to \ell \ell$ production, where $\ell = e$,$\mu$.  The
selected lepton must have a transverse momentum $\pt >30 \GeV$ and pass the ``medium''
identification requirement.

The $\tauhadvis$ candidate is required to have $\pt > 25 \GeV$, pass the ``medium'' BDT-based
identification requirement and lie in the range $|\eta|<2.3$.  The latter requirement is
motivated by a larger rate of electrons misidentified as $\tauhadvis$ candidates  at higher $|\eta|$
values: the rate is above 10\% for $|\eta|>2.3$, while it ranges from 0.5\% to 3\% for lower $|\eta|$ values.
If there is more than one $\tauhadvis$ candidate, the candidate with the highest $\pt$
is selected and the others are treated as jets.  Finally, the identified lepton and the $\tauhadvis$
are required to have opposite electric charge.

Subsequently, the following selection requirements are applied:

\begin{itemize}
  \item $\Delta \phi(\tauhadvis, \ell) > 2.4$.
  \item $\mt(\ell,\met) \equiv \sqrt{2\pt(\ell)\met \big[1-\cos\Delta\phi(\ell,\met)\big]} < 40 \GeV$.
  \item For the $\tau_{e}\tauhad$ channel, events are vetoed if the
    invariant mass of the electron and the visible $\tau$ lepton decay
    products is in the range $80<m_{\text{vis}}(e,\tauhadvis)<110 \GeV$.
\end{itemize} 

The requirement on $\Delta \phi(\tauhadvis, \ell)$ gives an overall reduction of SM backgrounds with
little signal loss. The requirement on $\mt(\ell,\met)$, the distribution of which is 
shown in Figure~\ref{fig:dphi}(a), serves to
remove events that originate from processes containing a $W$ boson: in signal events, the missing
transverse momentum is usually in the same direction as the $\taulep$, resulting in a low value of
$\mt(\ell,\met)$.  The requirement on $m_{\text{vis}}(e,\tauhadvis)$ reduces the contribution of
$Z\rightarrow ee$ decays, where an electron is misidentified as a $\tauhadvis$ candidate.  These
selection criteria define the inclusive $\taulep\tauhad$ selection.

\subsection{$\tauhad\tauhad$ channel}\label{sec:hadhad_selection}

Events in the $\tauhad\tauhad$ channel are selected by a trigger that requires a single $\tauhadvis$
satisfying the ``medium'' $\tauhadvis$ identification criterion  with $\pt>80$~$\GeV$.  The leading $\tauhadvis$ candidate in
$\pt$ must geometrically match the HLT object.  A $\pt$ requirement is applied to the leading
$\tauhadvis$ candidate, $\pt>110$~$\GeV$, and to the sub-leading $\tauhadvis$ candidate, $\pt>55$~$\GeV$.
Furthermore, the leading (sub-leading) $\tauhadvis$ candidate has to satisfy the ``medium'' (``loose'')
$\tauhadvis$ identification criterion.  Events with electrons or muons fulfilling the loose selection criteria described in
Section~\ref{sec:reconstruction} (with the exception of the isolation requirement) are vetoed to reduce
electroweak background processes and guarantee orthogonality with the $\taulep\tauhad$ channel.

The leading and sub-leading $\tauhadvis$ candidates must have opposite electric charge and have
a back-to-back topology in the transverse plane,
$\Delta\phi(\tau_{\mathrm{had}\mhyphen\mathrm{vis},1},
\tau_{\mathrm{had}\mhyphen\mathrm{vis},2})>2.7$. The distribution of
$\Delta\phi(\tau_{\mathrm{had}\mhyphen\mathrm{vis},1}, \tau_{\mathrm{had}\mhyphen\mathrm{vis},2})$
before this requirement is shown in Figure~\ref{fig:dphi}(b).  This selection defines the inclusive
$\tauhad\tauhad$ selection.

\begin{figure}[tbh]
  \centering
  \includegraphics[width=0.45\textwidth]{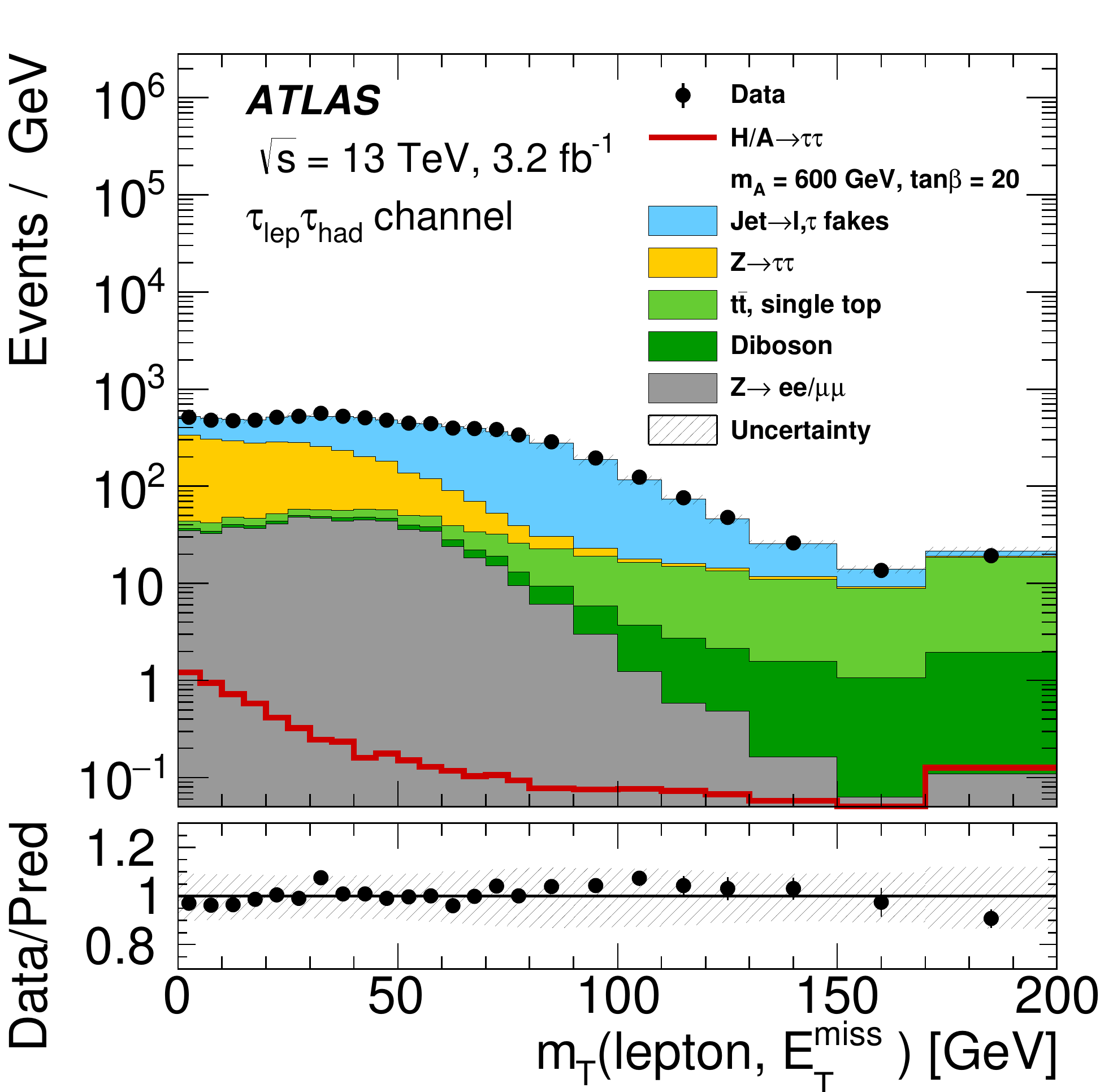}
  \includegraphics[width=0.45\textwidth]{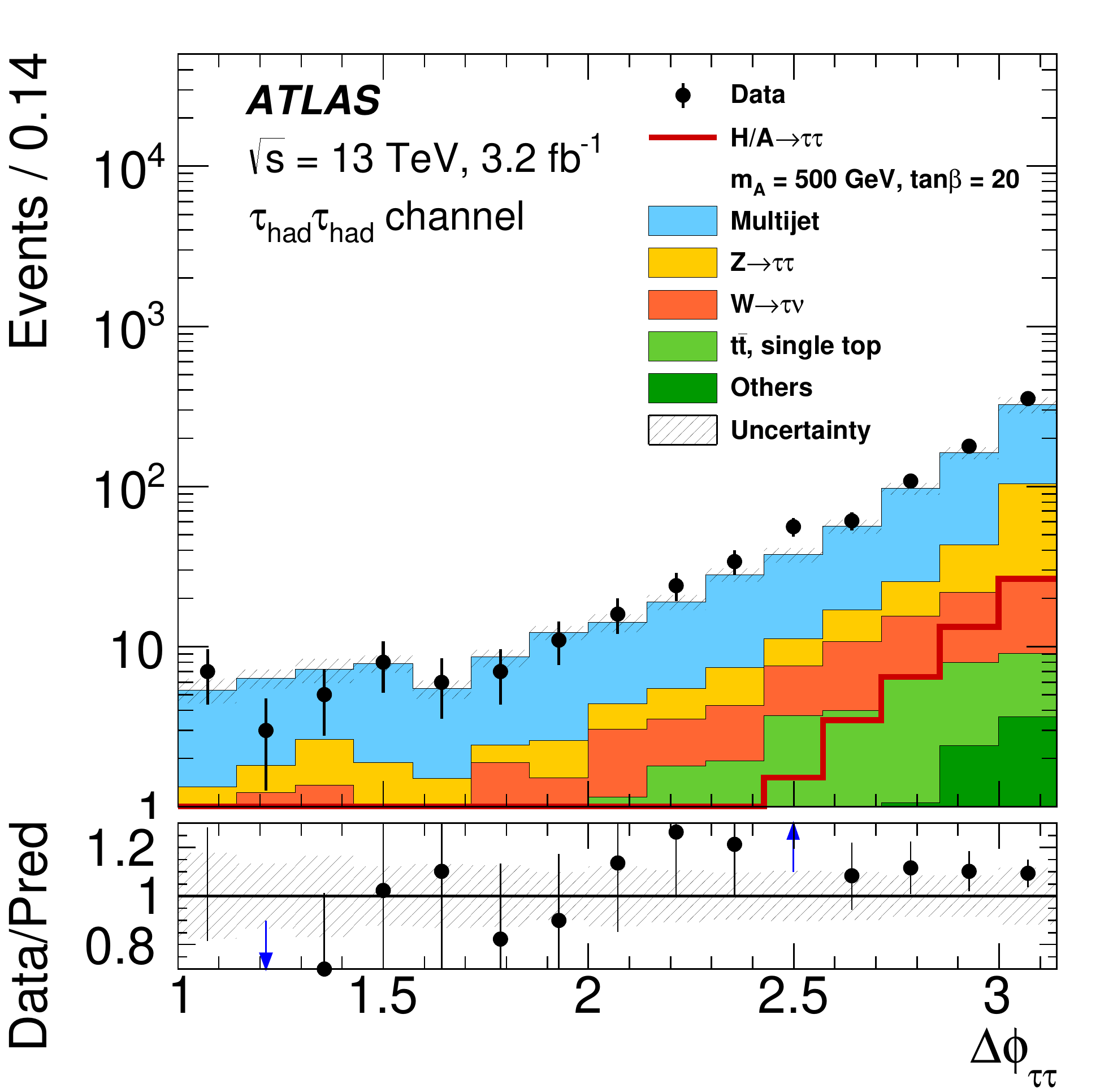}
  \caption{The distributions of (a) $\mt(\ell,\met)$ in the $\taulep\tauhad$ channel and (b)
    $\Delta\phi(\tau_{\mathrm{had}\mhyphen\mathrm{vis},1},
    \tau_{\mathrm{had}\mhyphen\mathrm{vis},2})$ for the $\tauhad\tauhad$ channel for the inclusive
    selection with the criterion for the variable displayed removed.  The label ``Others'' in (b)
    refers to contributions due to diboson, $Z(\to\ell\ell)$+jets and $W(\to\ell\nu)$+jets production.
    Bins have a varying size and overflows are included in the last bin of the
    distribution on the left.
  }
\label{fig:dphi}
\end{figure}

\subsection{Event categories} \label{sec:categories}
In the search for \Zprime{} bosons, the event selections described in
Sections~\ref{sec:lephad_selection} and~\ref{sec:hadhad_selection} result in a signal selection
efficiency\footnote{The term ``signal selection efficiency'' refers to the fraction of signal
  events decaying to $\taulep\tauhad$ or $\tauhad\tauhad$ that are subsequently reconstructed within
  the detector acceptance and pass the selection requirements.}  varying between 0.8\% (2.0\%) at
$m_{\Zprime}=500 \GeV$ and 3.4\% (3.8\%) at $m_{\Zprime}=2.5 \TeV$ for the $\taulep\thad$
($\thad\thad$) channel.  In the search for the $H/A$ bosons, events satisfying the inclusive
selection are categorized to exploit the two different signal production modes as follows:
\begin{itemize}
  \item $b$-veto: no  $b$-tag jets in the event,
  \item $b$-tag:  at least one  $b$-tag jet in the event.
\end{itemize}
In the \bveto{} category, the $H/A$ signal selection efficiency varies
between 2\% at $m_A=200 \GeV$ and 7\% at $m_A=1.2 \TeV$ for the gluon--gluon fusion production
mode in the $\taulep\thad$ channel, and from 0.1\% to 15\% in the $\thad\thad$ channel
for the same mass range.
In the \btag{} category, the signal selection efficiency varies
from 0.5\% at $m_A=200~\GeV$ to 2\% at $m_A=1.2~\TeV$ for the $b$-associated production mode 
in the $\taulep\thad$ channel, and from 0.1\% to 6\% in the $\thad\thad$ channel. 

\subsection{Di-tau mass reconstruction} \label{sec:mass}
The di-tau mass reconstruction is critical in achieving good separation between 
signal and background. However, its reconstruction is challenging due to the presence of
neutrinos from the $\tau$ lepton decays.
The mass reconstruction used for both the $\tauhad\tauhad$ and $\taulep\tauhad$ channels 
is the total transverse mass, defined as
\begin{equation}
\label{eq:mttot}
\mt^{\text{tot}} = \sqrt{ \mt^{2}(\MET, \tau_1) + \mt^{2}(\MET, \tau_2) +  \mt^{2}(\tau_1, \tau_2)}, 
\end{equation}
  where  $\mt(a, b)$ is defined as
\begin{equation}
  \mt(a, b) = \sqrt{2\pt(a) \pt(b) [1-\cos\Delta\phi (a, b)]} 
  \end{equation}
and $\tau$ refers to the visible decay of the $\tau$ lepton ($\ell$ or $\tauhadvis$).
More complex mass reconstruction techniques 
were investigated, but they did not improve the expected sensitivity.

\section{Background estimation}
\label{sec:background}
The background processes can be categorized according to whether the electron/muon and/or the
$\tauhadvis$ are correctly identified. Backgrounds from processes with correctly identified
$\tauhadvis$, electrons and muons, or where the $\tauhadvis$ is due to a misidentified electron/muon
in the $\taulep\tauhad$ channel,  are estimated from simulation.  
Data-driven techniques are used for processes where the $\tauhadvis$, or
both the lepton and $\tauhadvis$ are misidentified.  
The background contributions originating from processes where only the 
lepton is misidentified are found to be negligible.

\begin{table}[htb]
\centering
  \footnotesize
\begin{tabular}{l l l}
\hline
\hline
$\taulep\tauhad$ signal region &  & $\Delta \phi(\tau, \ell) > 2.4$, $\mt(\ell,\met) < 40~\GeV$, \\
                           &  & Veto $ 80 < m_{e,\tau} < 110~\GeV$ for $\tau_e\tauhad$, \\
                           &  & $N_{b\textrm{-tag}}\ge 1$ ($b$-tag category), $N_{b\textrm{-tag}}= 0$ ($b$-veto category) or \\
                           &  & no requirement ($\Zprime$ category)\\
\hline
$W$+jets$/t\bar{t}$ fake factor control region &  & $\mt(\ell,\met) >$ 70 (60)~\GeV{} for $\tau_e\tauhad$ ($\tau_\mu\tauhad$),\\
                                                     &  & different $\tauhadvis$ identification for anti-$\tauhad$ sub-region\\
\hline
$t\bar{t}$ validation region                     &  & $N_{b\textrm{-tag}}\ge 1$, $\mt(\ell,\met) > 100~\GeV$\\
\hline
Multi-jet fake factor control region     &  &  invert $e,\mu$ isolation requirement, \\
&                         &  different $\tauhadvis$ identification for anti-$\tauhad$ sub-region \\
\hline
Multi-jet control region for                 &   & $\mt(\ell,\met) <$ 30~\GeV, no $e,\mu$ isolation requirement, \\
$r_{\text{MJ}}$ estimation                    &   & no $\tauhadvis$ passing loose identification, \\
                                             &   & $N_{\textrm{jet}}\ge 1$ ($b$-veto category), $N_{\textrm{jet}} \ge 2$ ($b$-tag category) \\
\hline 
Control region for correction of electrons  &   & $ 80 < m_{e,\tau} < 110~\GeV$ for 1-track $\tauhadvis$ \\
misidentified as $\tauhadvis$               &   & $ 90 < m_{e,\tau} < 100~\GeV$ for 3-track $\tauhadvis$ \\
\hline\hline
 & & \\
\hline\hline
$\tauhad\tauhad$ selection &  & $\Delta \phi({\tauhadvis}_{,1}, {\tauhadvis}_{,2}) > 2.7$, \\
&  & $N_{b\textrm{-tag}}\ge 1$ ($b$-tag category), $N_{b\textrm{-tag}}= 0$ ($b$-veto category) or \\
                                     & & no requirement ($\Zprime$ category)\\
\hline
Multi-jet fake factor control region & & pass single-jet trigger, leading $\tauhadvis$ fails medium identification, \\
                                     & & no tracks, nor charge requirements for leading $\tauhadvis$, \\
                                     & & $\frac{\pt^{{\tauhadvis}_{,2}}}{\pt^{{\tauhadvis}_{,1}}} > 0.3$, no $\Delta\phi({\tauhadvis}_{,1}, {\tauhadvis}_{,2})$ requirement \\
\hline
Fake-rate control region             & & pass single-muon trigger,  isolated muon  with $\pt > 55~\GeV$, \\
                                     & & $\tauhadvis$ with $\pt > 50~\GeV$, $\Delta\phi(\mu,\tauhadvis)>2.4$, \\
                                     & &  $\sum_{L=\mu,\tau}\cos\Delta\phi(L,\met)<0$ (for $b$-veto category only) \\
\hline
\Wmunu{} control region for          & & pass single-muon trigger,  isolated muon  with $\pt > 110~\GeV$, \\
$W$+jets \MT{} correction            & & $\tauhadvis$ with $\pt > 55~\GeV$ \\
\hline
\hline
\end{tabular}
    \caption{Description of the control regions used in the $\taulep\tauhad$ and $\tauhad\tauhad$ channels.}
    \label{tab:control-regions}
\end{table}

\subsection{$\taulep\tauhad$ background estimate}
\label{sec:lephadbkg}

The main backgrounds in the $\taulep\tauhad$ channel arise from $Z \to \tau\tau$ production,
followed by processes with a misidentified $\tauhadvis$ in the $b$-veto category and $t\bar{t}$
production, with either a true $\tau$ lepton or a jet misidentified as a $\tauhadvis$, in the
$b$-tag category.

Background processes where the $\tauhad$ candidate, or both the lepton and $\tauhad$ candidates,
arise from misidentified jets are dominated by $W$+jets ($t\bar{t}$) and multi-jet processes, for
the $b$-veto ($b$-tag) category.  A data-driven ``fake factor'' (FF) technique is used to estimate
the contribution of these processes to the signal region.  The fake factors are derived separately
for the $b$-veto and $b$-tag categories using fake factor control regions (see
Table~\ref{tab:control-regions}) dominated by a particular background process (Pr), and are defined
as:

\begin{equation}\label{eq:wjets-fake-factor}
\text{FF}(\text{Pr}) = \frac{N(\text{nominal $\tauhadvis$ ID}, \text{Pr})}{N(\text{anti-$\tauhadvis$ ID}, \text{Pr})},
\end{equation}

where $N(\text{nominal $\tauhadvis$ ID}, \text{Pr})$ is the number of $\tauhadvis$ candidates in data
satisfying the ``medium'' $\tauhadvis$ identification criterion and
$N(\text{anti-$\tauhadvis$ ID}, \text{Pr})$ is the number of $\tauhadvis$ candidates failing this criterion
but meeting a loose requirement on the BDT score. The latter requirement defines the
``anti-$\tauhad$'' sub-region, which selects the same kind of objects mimicking $\tauhadvis$ candidates
as those fulfilling the identification criteria.  The true $\tauhad$ contamination in
the fake factor control regions is subtracted using simulation.
In all the control regions, the fake factors are parameterized as a function of the transverse
momentum and number of tracks of the reconstructed $\tauhadvis$ object.

The fake factor for $W$+jets and $t\bar{t}$ backgrounds, FF$(W$+jets$/t\bar{t})$, is measured in a
fake factor control region that is identical to the signal region, except that the $\mt(\ell,\met)$
selection criterion is reversed to $\mt(\ell,\met)> 60\,(70)~\GeV$ for the $\tau_{\mu}\tauhad$
($\tau_{e}\tauhad$) channel.  The purity of the $W$+jets background in the $b$-veto category of the
control region is about $ 95\%$, while in the $b$-tag category both the $W$+jets and $t\bar{t}$
processes contribute.  The fake factor value for the $b$-tag category was found to be compatible
with the value corresponding to the $b$-veto category. To improve the statistical precision, the
fake factor measured in a control region without requirements on the number of $b$-tags is used for
the $b$-tag category.  The same fake factor is used in the search for the \Zprime{} boson.  For
multi-jet events (MJ), the fake factor FF$(\text{MJ})$ is measured in a fake factor control region
defined by inverting the isolation requirement on the electron or muon.  The purity of multi-jet
events in this control region exceeds $99$\%. The fake factors are derived separately for the $b$-veto
and $b$-tag categories by requiring no $b$-tag and at least one $b$-tag, respectively. For the
$Z^{\prime}$ analysis, no $b$-tag requirement is considered.

The shapes and normalization of background contributions in the signal region are then estimated by
applying these fake factors to events that pass the anti-$\tauhad$ region selection but
otherwise satisfy all signal region requirements.  In this analysis, the fake factors are combined
and weighted by the predicted contribution of each background process to the anti-$\tauhad$ region:

\begin{equation}\label{eq:combined-fake-factor}
\text{FF}(\text{comb}) = \text{FF}(W+\text{jets}/t\bar{t}) \times r_{W/t\bar{t}} + \text{FF}(\text{MJ}) \times r_{\text{MJ}},
\end{equation}

where $r_{\text{MJ}}$ denotes the fraction of multi-jet events in the anti-$\tauhad$ region and
$r_{W/t\bar{t}} = 1 - r_{\text{MJ}}$. This neglects the differences between the fake factors for 
$W$+jets/$t\bar{t}$ and other processes, such as $Z$ production.  
The parameter $r_{\text{MJ}}$ is estimated, separately for the
$b$-veto and $b$-tag categories, in two steps using a data-driven method.  First, the rates at which jets are
misidentified as electrons or muons are measured from the ratio of leptons passing and failing the
lepton isolation requirement in a region enriched in  multi-jet events.  This multi-jet control region is defined 
in Table~\ref{tab:control-regions}. The predicted multi-jet rate is then applied to
events in the anti-$\tauhad$ sub-region that also fail the lepton isolation, in order to calculate
$r_{\text{MJ}}$ as a function of $\tauhadvis$ $\pt$ separately for the $\tau_e\tauhad$ and 
$\tau_\mu\tauhad$ channels.
When the fake factor is applied to the anti-$\tauhad$ sub-region events, the contributions from
correctly identified \tauhadvis{}  and from electrons and muons misidentified as 
\tauhadvis{} candidates are subtracted using the default MC simulation described in Section~\ref{sec:samples}.

Background processes where the electron or the muon is identified as a $\tauhadvis$ object are modelled using
simulation. The main source of such backgrounds is $Z (\rightarrow ee)$+jets events in the
$\tau_{e}\tauhad$ channel, which are reduced using the $m_{\text{vis}}(e,\tauhad)$ mass-window veto
described in Section~\ref{sec:lephad_selection}.  To account for mismodelling of electrons misidentified as
$\tauhadvis$ objects in $Z\rightarrow ee$+jets events, the simulation is corrected as a function of the
lepton $\eta$ using  data control regions defined by reversing the mass-window criterion, as listed in
Table~\ref{tab:control-regions}. The correction amounts to 15\% for three-track $\tauhadvis$, while for the one-track  $\tauhadvis$ objects the correction varies from 20\% in the barrel region to up to 200\% in the end-cap region.

The $\mttot$  distributions in the $\taulep\tauhad$ channel are shown in
Figures~\ref{fig:BGCR}(a) and \ref{fig:BGCR}(b) for the $W$+jets control region and the $t\bar{t}$ validation region,
respectively.  The latter is identical to the $b$-tag category definition, except for the $\mt(\ell,\met)$
requirement, which is reversed to  $\mt(\ell,\met) > 100 \GeV$.

\begin{figure}[]
  \centering

  \subfloat[]{\includegraphics[width=0.50\textwidth]{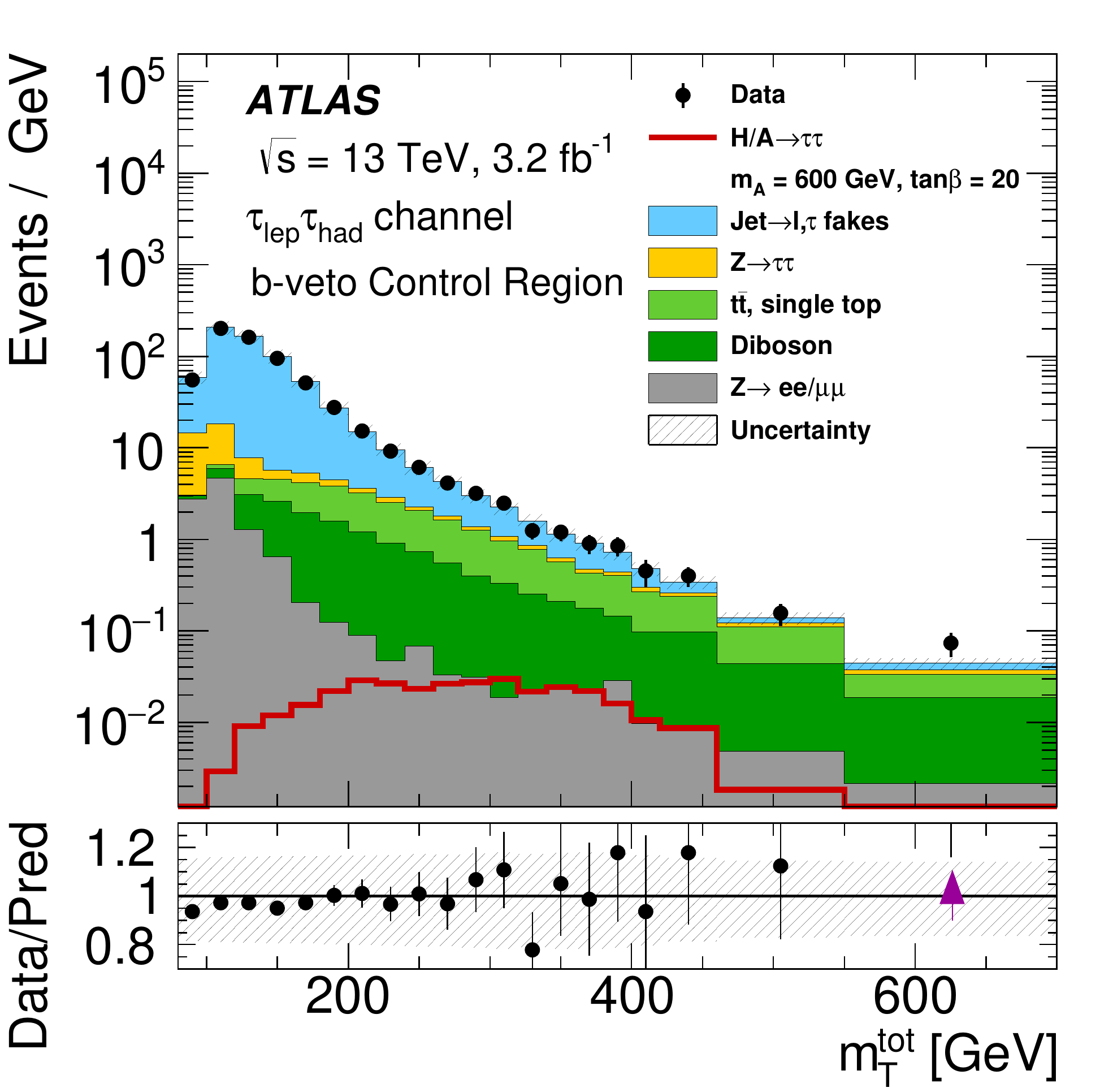} }
  \subfloat[]{\includegraphics[width=0.50\textwidth]{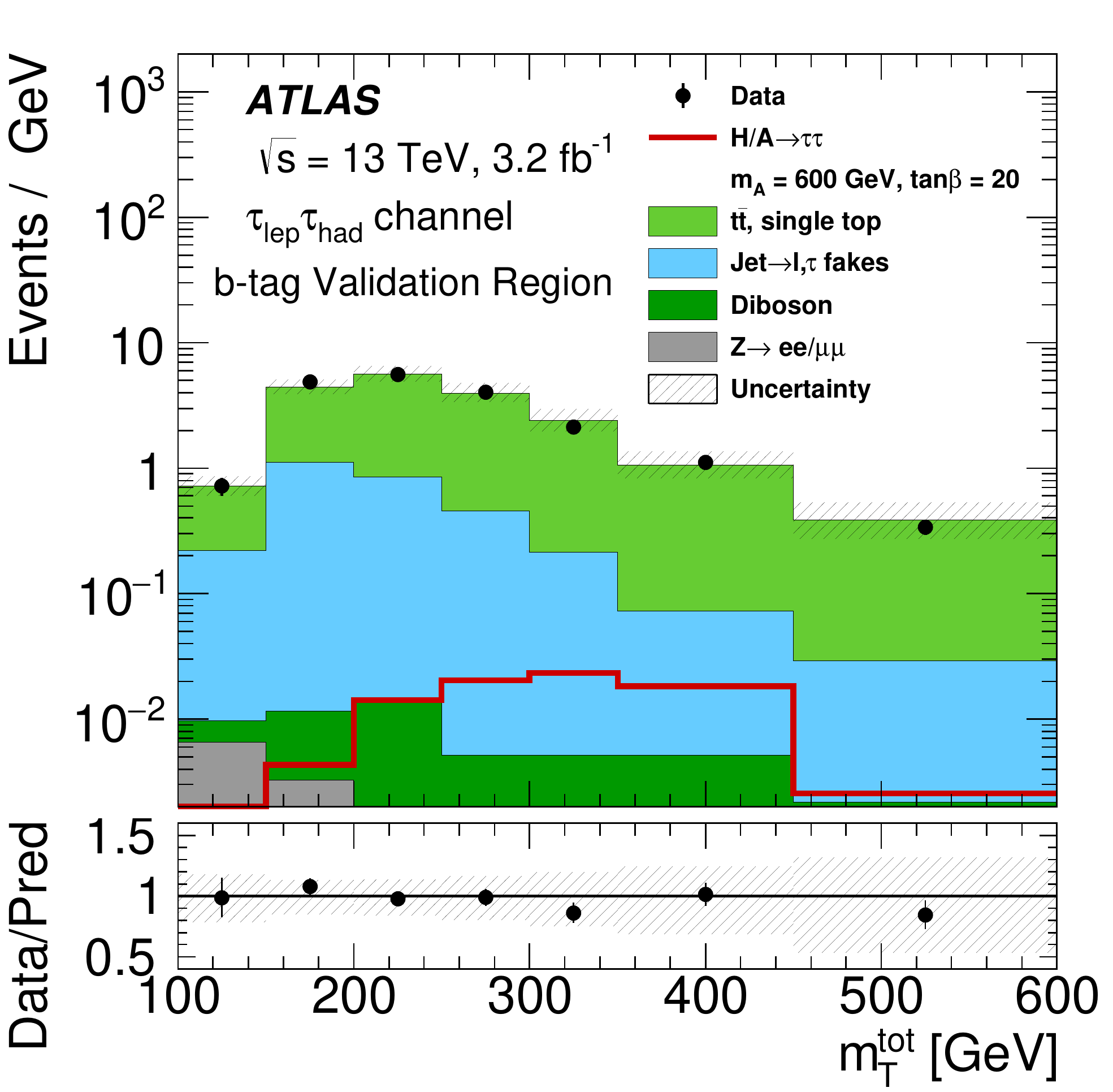} } \\
  \subfloat[]{\includegraphics[width=0.50\textwidth]{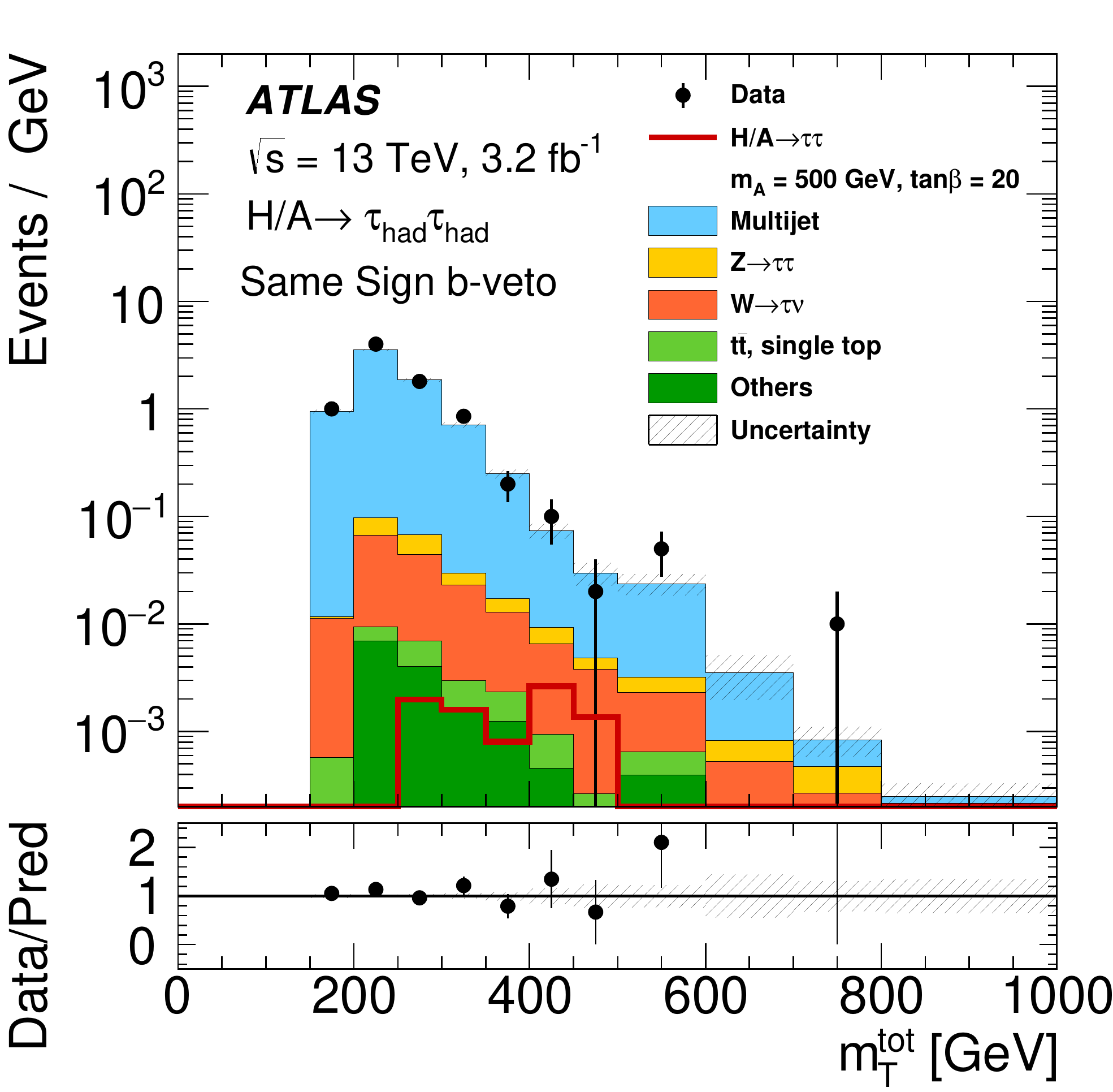} }
  \subfloat[]{\includegraphics[width=0.50\textwidth]{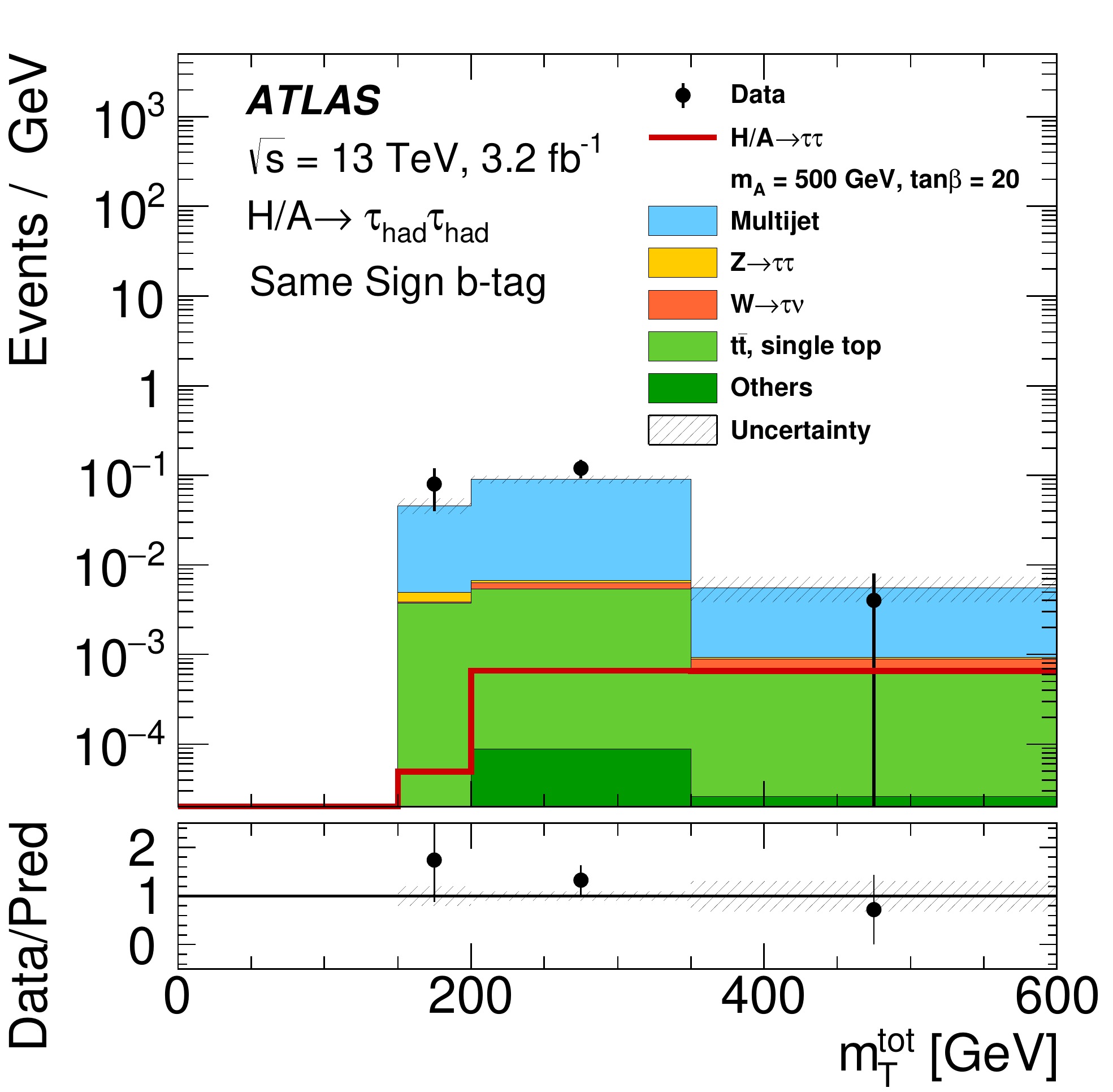} }
  
  \caption{The distributions of $\mttot$ in (a) the $\taulep\tauhad$ channel $W$+jets control
    region, (b) the $t\bar{t}$ validation region of the $\taulep\tauhad$ channel, (c) the
    $\tauhad\tauhad$ channel $b$-veto same-sign validation region and (d) the $\tauhad\tauhad$
    channel $b$-tag same-sign validation region. The various control and validation regions are
    defined in Table~\ref{tab:control-regions}.  The data are compared to the background prediction
    and a hypothetical MSSM $H/A\to\tau\tau$ signal ($m_A = 500\GeV$ and $\tan\beta = 20$).
    The Monte Carlo statistics of the signal is limited in the background-dominated regions. The
    label ``Others'' in (c) and (d) refers to contributions due to diboson, $Z(\to\ell\ell)$+jets and
    $W(\to\ell\nu)$+jets production.  The background uncertainty includes statistical and systematic
    uncertainties. The bins have a varying size and overflows are included in the last bin of the
    distributions. }

\label{fig:BGCR}
\end{figure}

\subsection{$\tauhad\tauhad$ background estimate}

The dominant background process for the $\tauhad\tauhad$ channel is multi-jet production, the cross
section of which is several orders of magnitude higher than that of the signal processes. Despite
the large suppression of this background thanks to the event selection, a sizeable contribution of
events with two jets misidentified as $\tauhadvis$ candidates remains. A fake factor technique is
used to normalize and model this background. Fake factors parameterized as a function of
$\pt(\tauhadvis)$ and the number of tracks are derived from a control region enriched with multi-jet
events, described in Table~\ref{tab:control-regions}. The factors are derived independently for the
$b$-tag and $b$-veto categories in the search for $H/A$ bosons, and inclusively in the search for
\Zprime{} bosons.  They are then applied to data events where the leading $\tauhadvis$ has passed
the $\tauhadvis$ identification requirement in the signal region, while the sub-leading $\tauhadvis$
candidate has passed only a loose requirement on the BDT score. The
contributions from non-multi-jet production processes are subtracted using simulation. 

For $t\bar{t}$ and $W$+jets events, along with other simulated background processes, the probability
of a jet being misidentified as a $\tauhadvis$ is modelled with a ``fake-rate'' technique.
The rates of jets being misidentified as a $\tauhadvis$ are measured
from data as a  function of the transverse momentum and number of tracks of the reconstructed \tauhadvis. 
The fake-rate control regions are described in Table~\ref{tab:control-regions} and are enriched 
in $W(\rightarrow \mu \nu)$+jets for the $b$-veto category and  $t\bar{t}$ events for the $b$-tag category, 

The fake rate is then applied
to the simulated events as a weight for each of the reconstructed $\tauhadvis$ that does not match
geometrically a true $\tau$ lepton. Fake rates derived in the fake-rate control region of the $b$-tag category are used for
simulated $t\bar{t}$ and single top quark events, while fake rates obtained in the $b$-veto control region 
are applied to the remaining processes. An additional weight is applied to $W\rightarrow \tau \nu$+jets 
events as a function of
\MT{}, to improve the modelling of the kinematics of the $W$+jets simulated events.  The reweighting function
is derived by fitting the ratio of the data to the simulation for the $\Wmunujets$ process in an
additional \Wmunu{} control region, defined in analogy with the inclusive signal 
selection and described in Table~\ref{tab:control-regions}.

A same-sign validation region, enriched with events where at least one jet is misidentified as a \tauhadvis,
is obtained by inverting the opposite-sign requirement of the two $\tauhadvis$ candidates. 
Distributions of $\mttot$ in the $\tauhad\tauhad$ channel same-sign validation region are shown in 
Figures~\ref{fig:BGCR}(c) and \ref{fig:BGCR}(d). The good performance multi-jet  background estimation method
is demonstrated by the agreement of the data with the background prediction.

\section{Systematic uncertainties}
\label{sec:systematics}
The signal efficiency and the background estimates are affected by uncertainties associated with
the detector simulation, the signal modelling and the data-driven background determination.

The integrated luminosity measurement has an uncertainty of $5\%$ 
and is used for all MC samples.  Uncertainties related to the detector are
included for all signal and backgrounds that are estimated using simulated samples. These uncertainties
are also taken into account for simulated samples that are used in the derivation of data-driven
background estimates. All instrumental systematic uncertainties arising from the reconstruction,
identification and energy scale of $\tauhadvis$ candidates, electrons, muons, ($b$-)jets and the
soft term of the \met{} measurement are considered. 
The effect of the energy scale uncertainties on the objects is propagated to the \met \:
determination.  
The electron, muon, jet and $\met$ systematic uncertainties described
above are found to have a very small effect.

Systematic uncertainties resulting from the data-driven background estimates are derived as
follows.  In the $\taulep\tauhad$ channel, the combined fake-factor method includes uncertainties in 
the $W$+jets$/t\bar{t}$ fake factors, the multi-jet fake factors, and the $r_{\text{MJ}}$ estimation.  For the
$W$+jets fake factors the main uncertainties arise from the dependence on
$\Delta\phi(\tauhadvis,\met)$, from the difference between the relative contributions of quark- and
gluon-initiated jets faking $\tauhadvis$ in the control region and the signal region, and from the
contamination by backgrounds other than $W$+jets in the control region, which are estimated using
simulation.  The uncertainty is parameterized as a function of the anti-$\tauhadvis$ \pt{} and amounts
approximately to $17\%$ for jets misidentified as one-track $\tau$ candidates and varies between
16\% and 34\%  for jets misidentified as three-track $\tau$ candidates.
Uncertainties related to non-$W$+jets events were studied and
have no significant impact on the fake-factor determination.
For the multi-jet fake factors and
$r_{\text{MJ}}$, the uncertainty is dominated by the number of data events in the
control region and the subtraction of the remaining non-multi-jet
backgrounds using simulation.  Typical values of the total uncertainties
for $r_{\text{MJ}}$ are between 7\% and 20\% and for the multi-jet fake factors between 10\% and 20\%,
depending on the channel and the $\tauhadvis$ candidate $\pt$.
In addition, the effect on the background estimate due to the 
anti-$\tauhad$ region definition is examined. The loose $\tauhadvis$ identification requirement
used in the definition of this region is varied to estimate the corresponding 
uncertainty, which is $5\%$ and $1\%$ in the $\tau_e\tauhad$
and the $\tau_\mu\tauhad$ channel, respectively.

In the $\tauhad\tauhad$ channel, the uncertainty in the fake-factor measurement used for the
multi-jet background estimation is obtained as the sum in quadrature of the statistical uncertainty
of the measurement and the difference between the fake factors determined from same-sign and from
opposite-sign events.  The fake rates for jets misidentified as $\tauhadvis$ are determined from
data. The main systematic uncertainty arises from the statistical uncertainty of the fake-rate
measurement and it ranges from 7\% to 30\% as a function of the $\tauhadvis$ $\pt$.
In the $\tauhad\tauhad$ channel, the uncertainty in the parameters of the function used to
reweight the $W\to\tau\nu$+jets background is propagated to the \MT{} distribution,
where its effect  ranges from 5\% to 20\%.

Theoretical cross-section uncertainties are considered for all backgrounds estimated using simulation.
For $Z$+jets and diboson production, uncertainties of $5\%$ and $6\%$ are
used, respectively, combining PDF$+\alphas$ and scale variation uncertainties in
quadrature. For $t\bar{t}$~\cite{Czakon:2011xx} and single top-quark~\cite{Aliev:2010zk, Kant:2014oha}
production, a $6\%$ uncertainty is assigned based on scale, PDF and top-quark mass uncertainties.
Additional uncertainties related to initial- and final-state radiation modelling, tune and
(for $\ttbar$ only) the choice of the \texttt{hdamp} parameter value in \texttt{POWHEG-BOX v2}, 
which controls the amount of radiation produced by the parton shower,  were also 
taken into account~\cite{ATLAS:TopMCsyst}.
The uncertainty in the fragmentation model is evaluated by comparing $\ttbar$ events generated
with \texttt{POWHEG-BOX v2} interfaced to either \texttt{HERWIG++}~\cite{Herwigpp} or
\texttt{PYTHIA6}. The \texttt{POWHEG+HERWIG++} and \texttt{aMC@NLO+HERWIG++} generators are used to
estimate the uncertainty in generating the hard scatter. 
The variation of the $b$-tag category acceptance for these uncertainties is from
 $-10\%$ to $+30\%$ ($-33\%$ to $+38\%$) in the $\taulep\tauhad$ ($\tauhad\tauhad$) channel.

Uncertainties related to signal modelling are discussed in the following.
Uncertainties due to the factorization and renormalization scale choices are estimated from the
effect on the signal acceptance of doubling or halving these factors either coherently or oppositely.
Uncertainties due to the  initial- and final-state radiation, as well as multiple parton interaction
for the signal are also taken into account. These uncertainties are estimated from the 
{\tt PYTHIA8} A14 tune~\cite{ATL-PHYS-PUB-2014-021} for the  $b$-associated production
and  the AZNLO {\tt PYTHIA8} tune~\cite{AZNLO:2014} for the gluon--gluon fusion production.
The envelope of the variations resulting from the use of the alternative PDFs in the
PDF4LHC15$\_$nlo$\_$100~\cite{Ball:2014uwa} set is used in order to
estimate the PDF uncertainty for gluon--gluon fusion production. For the  $b$-associated production
uncertainty, a comparison among NNPDF30\_nlo\_as\_0118\_nf\_4~\cite{Ball:2014uwa}, 
CT14nlo\_NF4~\cite{Dulat:2015mca}, MSTW2008nlo68cl\_nf4~\cite{Martin:2010db} 
and CT10\_nlo\_nf4~\cite{Lai:2010vv} PDF sets is employed.
Since no statistically significant effect on the shape of the reconstructed mass distribution is observed, each
contribution is taken solely as a normalization uncertainty. The total uncertainty ranges between $15\%$ and $25\%$.

\section{Results}
\label{sec:results}
The parameter of interest is the signal strength, $\mu$.  It is defined as the ratio
of the observed to the predicted value of the cross section times branching fraction, where the
prediction is evaluated for a particular MSSM or $\Zprime$ assumption. Hence, the value $\mu=0$
corresponds to the absence of a signal, whereas the value $\mu=1$ indicates the presence of a signal
as predicted by the theoretical model under study.  To estimate $\mu$, a likelihood function
constructed as the product of Poisson probability terms is used. Signal and background predictions
depend on systematic uncertainties, which are parameterized as nuisance parameters and are
constrained using Gaussian probability distributions.  For the MSSM Higgs boson search a binned
likelihood function is constructed in bins of the $\MT$ distributions, chosen to ensure sufficient
background statistics in each bin.  The search for a \Zprime{} boson is a counting experiment, summing the
number of events above a certain \mttot{} threshold.  The threshold is chosen for each \Zprime{}
mass hypothesis to maximize the expected significance and ranges from 400~\GeV\ at low \Zprime{} mass to
750~\GeV\ at high \Zprime{} mass.  The asymptotic approximation is used with the test statistic
$\tilde{q}_\mu$ \cite{Cowan:2010js} to test the compatibility of the data with the assumed signal.

The number of observed $\taulep\tauhad$ and $\tauhad\tauhad$ data events, along with the predicted
event yields from background and signal processes, in the signal regions are shown in Table
\ref{tab:data-pred-comparison}. The observed event yields are compatible with the expected event
yield from SM processes, within uncertainties. The $\mttot$ mass distributions are shown in
Figure~\ref{fig:mttotPF}. The results from the $\taulep\tauhad$ and $\tauhad\tauhad$ channels are
combined to improve the sensitivity to $H/A$ and \Zprime{} boson production.

\begin{table}[t]
\begin{center}

  \begin{tabular}{ l rcl rcl}
\multicolumn{7}{c}{$\tau_{e}\tauhad$ channel} \\
\hline
                                     &  \multicolumn{3}{c}{$b$-tag category}  &  \multicolumn{3}{c}{$b$-veto category} \\ 
\hline
$Z\to\tau\tau$+jets                  &  42    &  $\pm$ &    7   &  4500   & $\pm$  &   250 \\
Jet $\to \ell, \tauhadvis$ fakes     &  128   &  $\pm$ &    18  &  5400   & $\pm$  &   350 \\
$Z\to\ell\ell$+jets                  &  3.6   &  $\pm$ &    1.5 &  590    & $\pm$  &   120 \\
$t\bar{t}$ and single top quark      &  115   &  $\pm$ &    16  &  35     & $\pm$  &   5   \\
Diboson                              &  0.33  &  $\pm$ &    0.07&  44     & $\pm$  &   4   \\
\hline                                                                                    
Total prediction                     &  289   &  $\pm$ &    24  &  10600  & $\pm$  & 360   \\
Data                                 &  275   &        &        &  10619  &        &       \\
\hline                                                                                    
$ggH$ $m_{A}=500 \GeV, \tan\beta=20$ & 0.020  &  $\pm$ &   0.010 &   1.2  & $\pm$  &  0.2  \\
$bbH$ $m_{A}=500 \GeV, \tan\beta=20$ & 6.4    &  $\pm$ &   1.7   &   7.4  & $\pm$  &  1.9  \\
\hline
 & &  \\
 & &  \\
\multicolumn{7}{c}{$\tau_{\mu}\tauhad$ channel} \\
\hline
                                     &  \multicolumn{3}{c}{$b$-tag category}  &  \multicolumn{3}{c}{$b$-veto category} \\ 
\hline
$Z\to\tau\tau$+jets                  &  42    &  $\pm$  &   6     &  5500  &  $\pm$ &   300\\
Jet $\to \ell, \tauhadvis$ fakes     &  109   &  $\pm$  &   14    &  2760  &  $\pm$ &   170\\
$Z\to\ell\ell$+jets                  &  5.2   &  $\pm$  &   0.6   &  830   &  $\pm$ &   50 \\
$t\bar{t}$ and single top quark      &  136   &  $\pm$  &   15    &  40    &  $\pm$ &    5 \\
Diboson                              &  0.34  &  $\pm$  &   0.07  &  55    &  $\pm$ &    5 \\
\hline                                                                                     
Total prediction                     &  293   &  $\pm$  &   19    &  9200  &  $\pm$ &  300 \\
Data                                 &  312   &         &         &  9163  &        &          \\
\hline                                                                                     
$ggH$ $m_{A}=500 \GeV, \tan\beta=20$ & 0.016  &  $\pm$  &  0.005  & 1.1    &  $\pm$ &   0.2    \\
$bbH$ $m_{A}=500 \GeV, \tan\beta=20$ & 3.3    &  $\pm$  &  1.3    & 6.4    &  $\pm$ &   1.7    \\
\hline

 & &  \\
 & &  \\
\multicolumn{7}{c}{$\tauhad\tauhad$ channel} \\
\hline
                                     &  \multicolumn{3}{c}{$b$-tag category}  &  \multicolumn{3}{c}{$b$-veto category} \\ 
\hline
$Z\to\tau\tau$+jets                  &  1.9   & $\pm$ &  0.3  &  146  &   $\pm$  & 20   \\
Multi-jet                            &  17    & $\pm$ &  3    &  396  &   $\pm$  & 16   \\
\Wtaunu + jets                       &  1.1   & $\pm$ &  0.2  &  45   &   $\pm$  & 7    \\
$t\bar{t}$ and single top quark      &  11    & $\pm$ &  3    &  4.5  &   $\pm$  & 0.9  \\
Others                               &  0.13  & $\pm$ &  0.03 &  6.3  &   $\pm$  & 0.8  \\
\hline                                                                                 
Total prediction                     &  31    & $\pm$ &  4    &  598  &   $\pm$  & 21   \\
Data                                 &  23    &       &       &  628  &          &      \\
\hline                                                                                 
$ggH$ $m_{A}=500 \GeV, \tan\beta=20$ & 0.034  & $\pm$ &  0.014& 2.2   &   $\pm$  &  0.7   \\
$bbH$ $m_{A}=500 \GeV, \tan\beta=20$ & 8      & $\pm$ &   3   & 15    &   $\pm$  &  5     \\
\hline    

  \end{tabular}

    \caption{Observed number of events and background predictions in the $b$-tag and $b$-veto
      categories for the $\tau_{e}\tauhad$, $\tau_{\mu}\tauhad$ and $\tauhad\tauhad$ channels. The
      background predictions and uncertainties are obtained from the statistical procedure discussed
      in Section~\ref{sec:results}. In the $\taulep\tauhad$ channel, the processes other than
      ``Jet $\to \ell, \tauhadvis$ fakes'' require a true hadronically decaying $\tau$ lepton or an electron or muon
      misidentified as a $\tauhadvis$.  The expected signal yields for the $m_{h}^{\text{mod}+}$
      scenario are shown for comparison.}
    \label{tab:data-pred-comparison}

\end{center}
\end{table}

\begin{figure}[tbh]
  \centering

  \subfloat[$\taulep\tauhad$ $b$-veto category]{\includegraphics[width=0.375\textwidth]{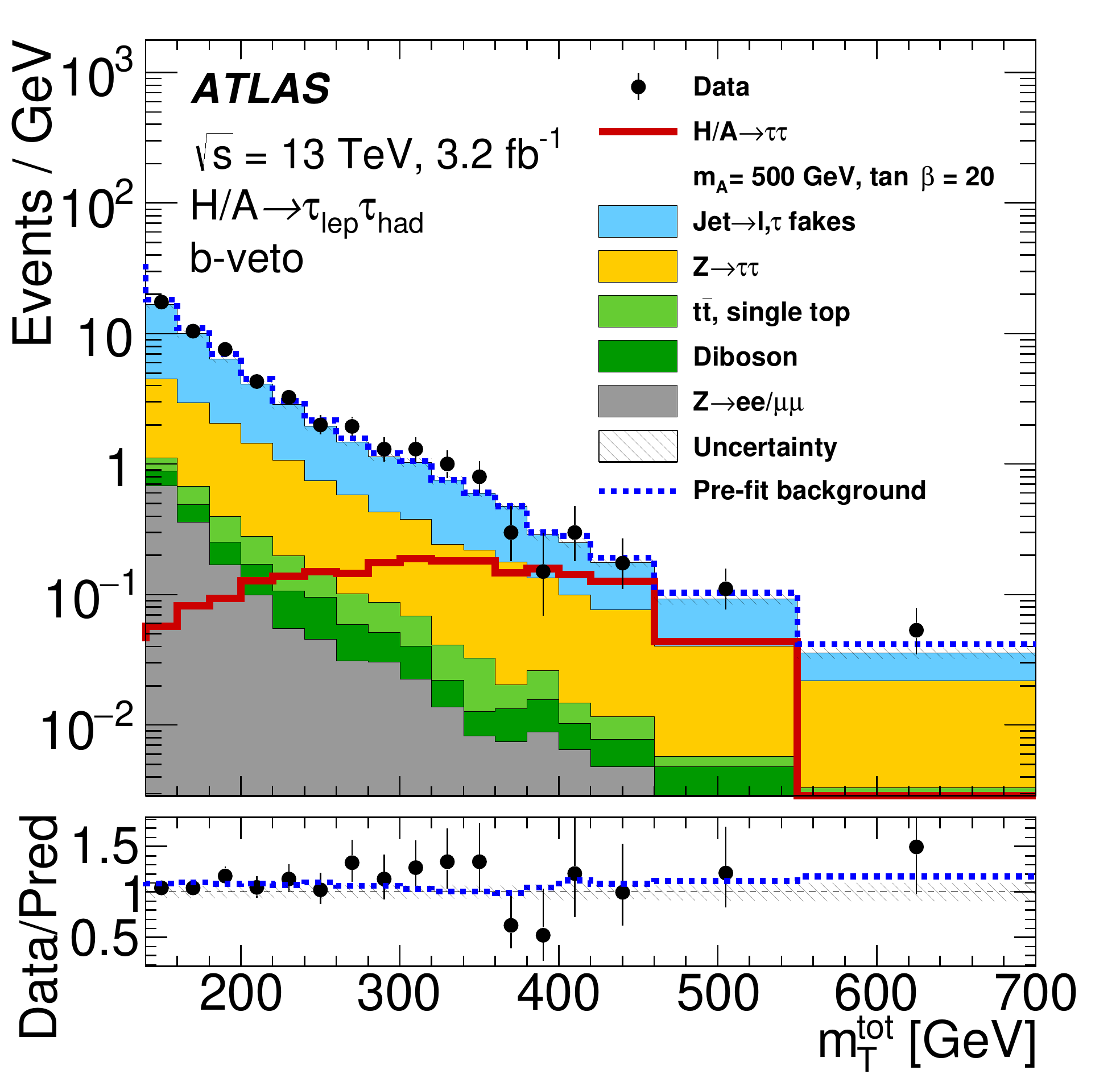} }
  \subfloat[$\tauhad\tauhad$ $b$-veto category]{\includegraphics[width=0.375\textwidth]{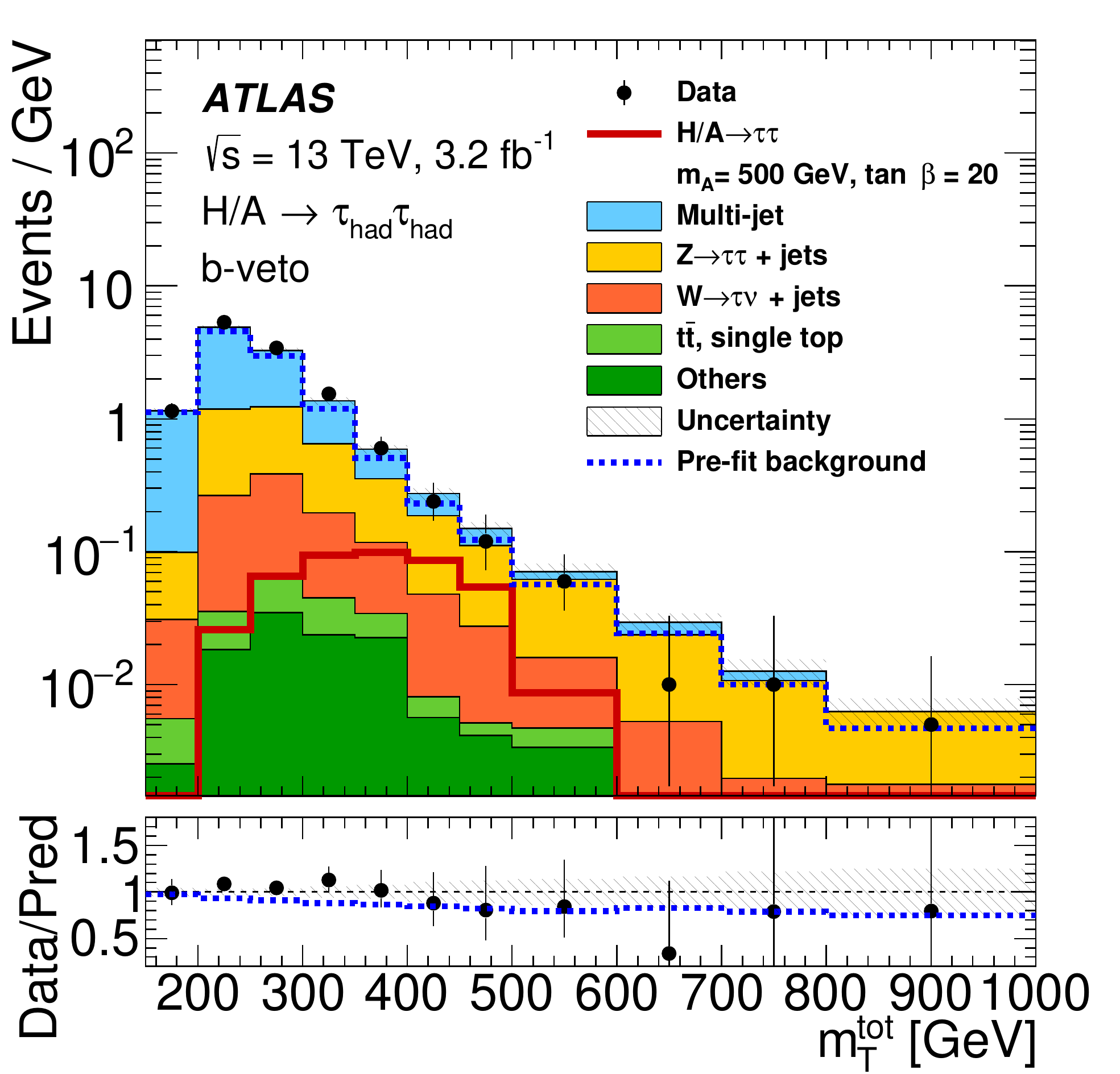} } \\
  \subfloat[$\taulep\tauhad$ $b$-tag category]{\includegraphics[width=0.375\textwidth]{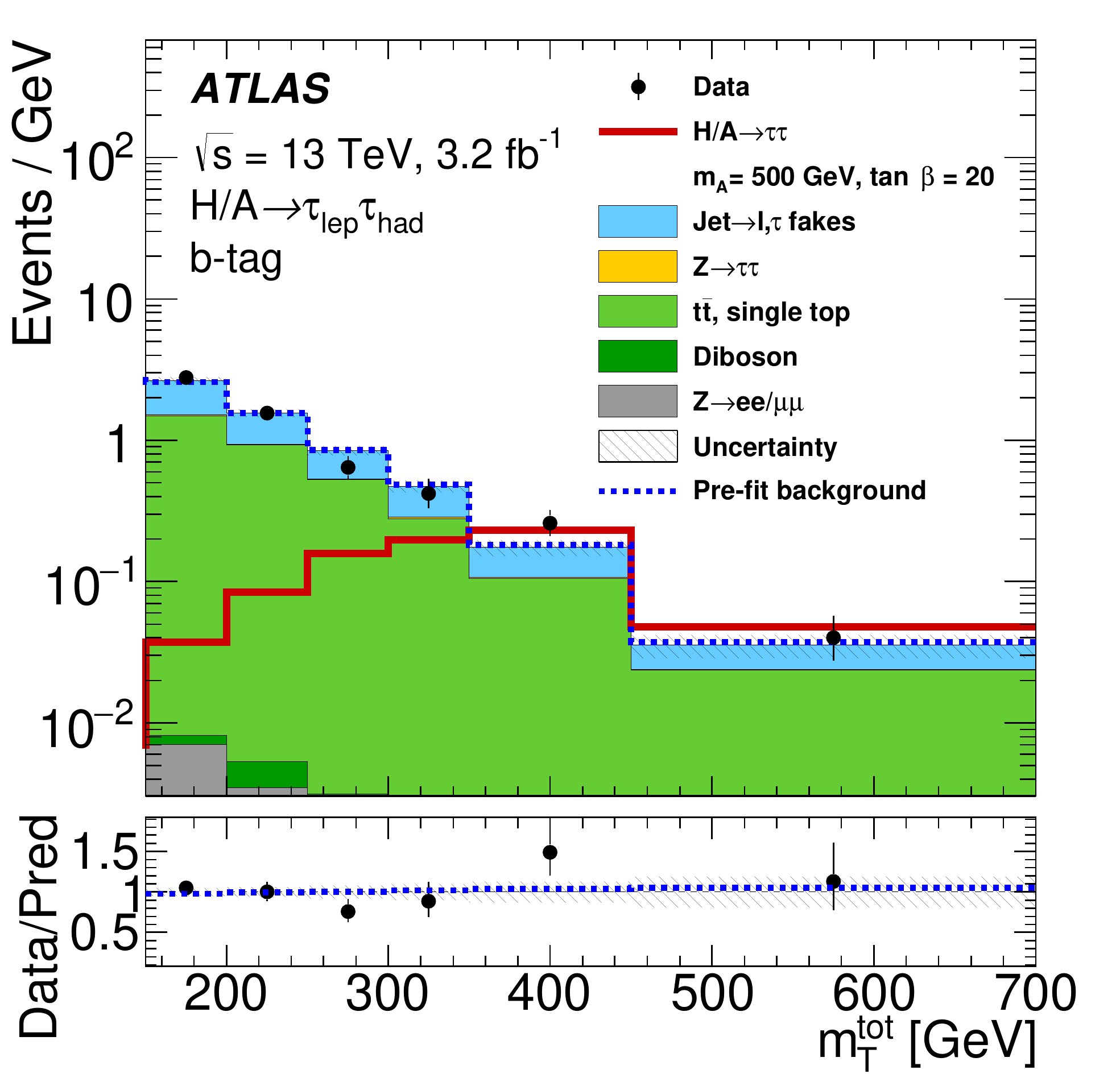} }
  \subfloat[$\tauhad\tauhad$ $b$-tag category]{\includegraphics[width=0.375\textwidth]{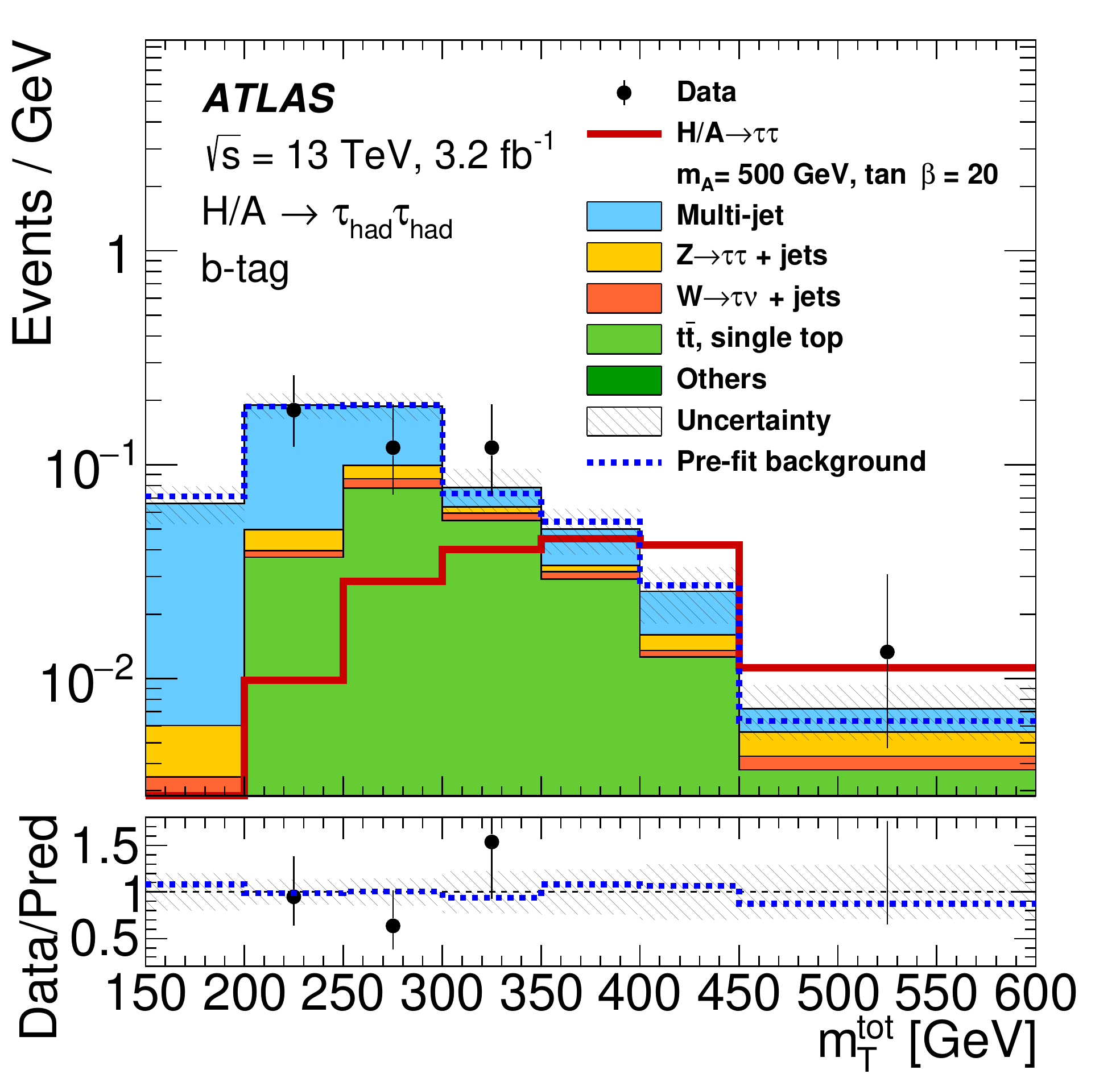} } \\
  \subfloat[$\taulep\tauhad$ inclusive category]{\includegraphics[width=0.375\textwidth]{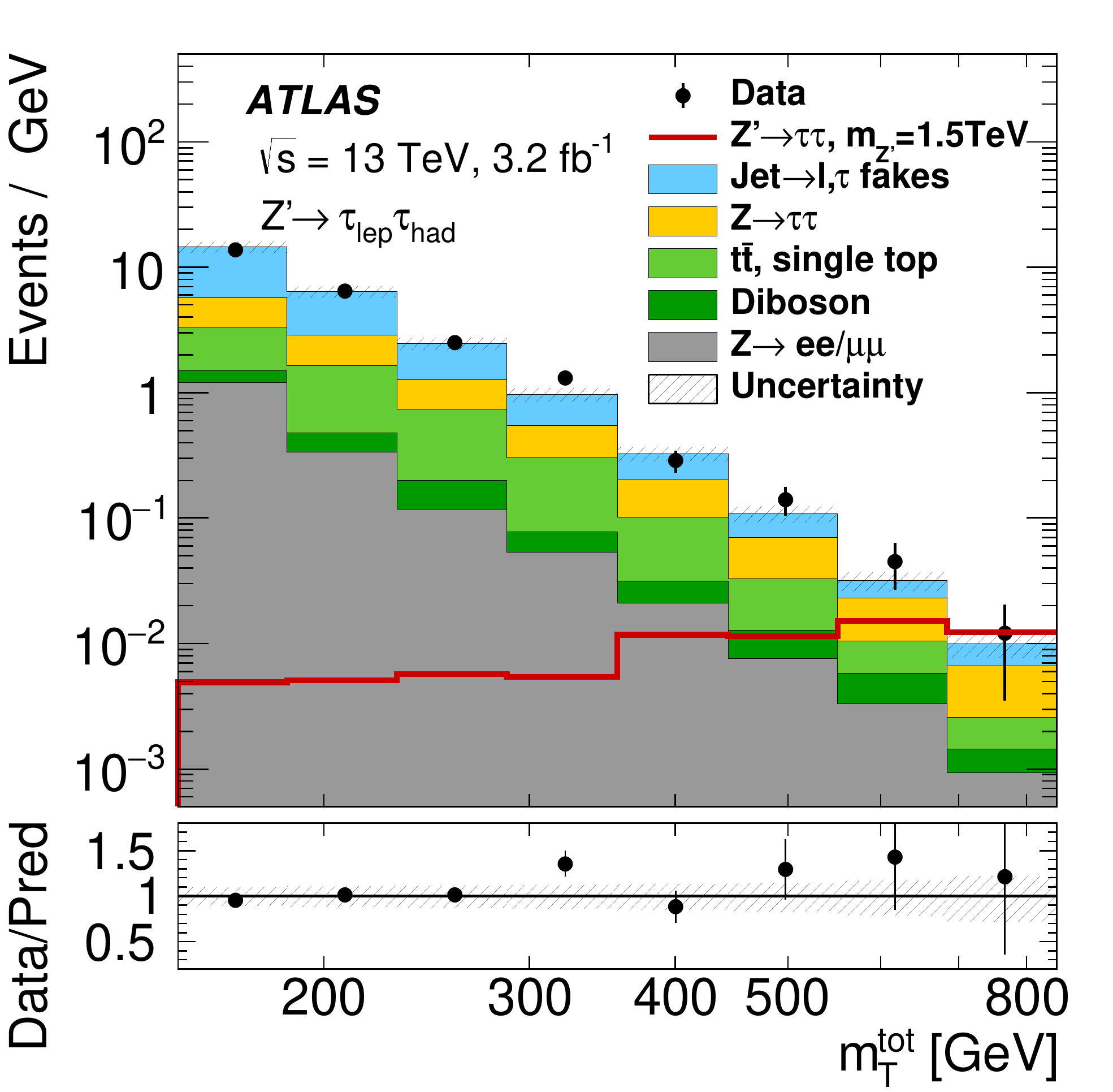} }
  \subfloat[$\tauhad\tauhad$ inclusive category]{\includegraphics[width=0.375\textwidth]{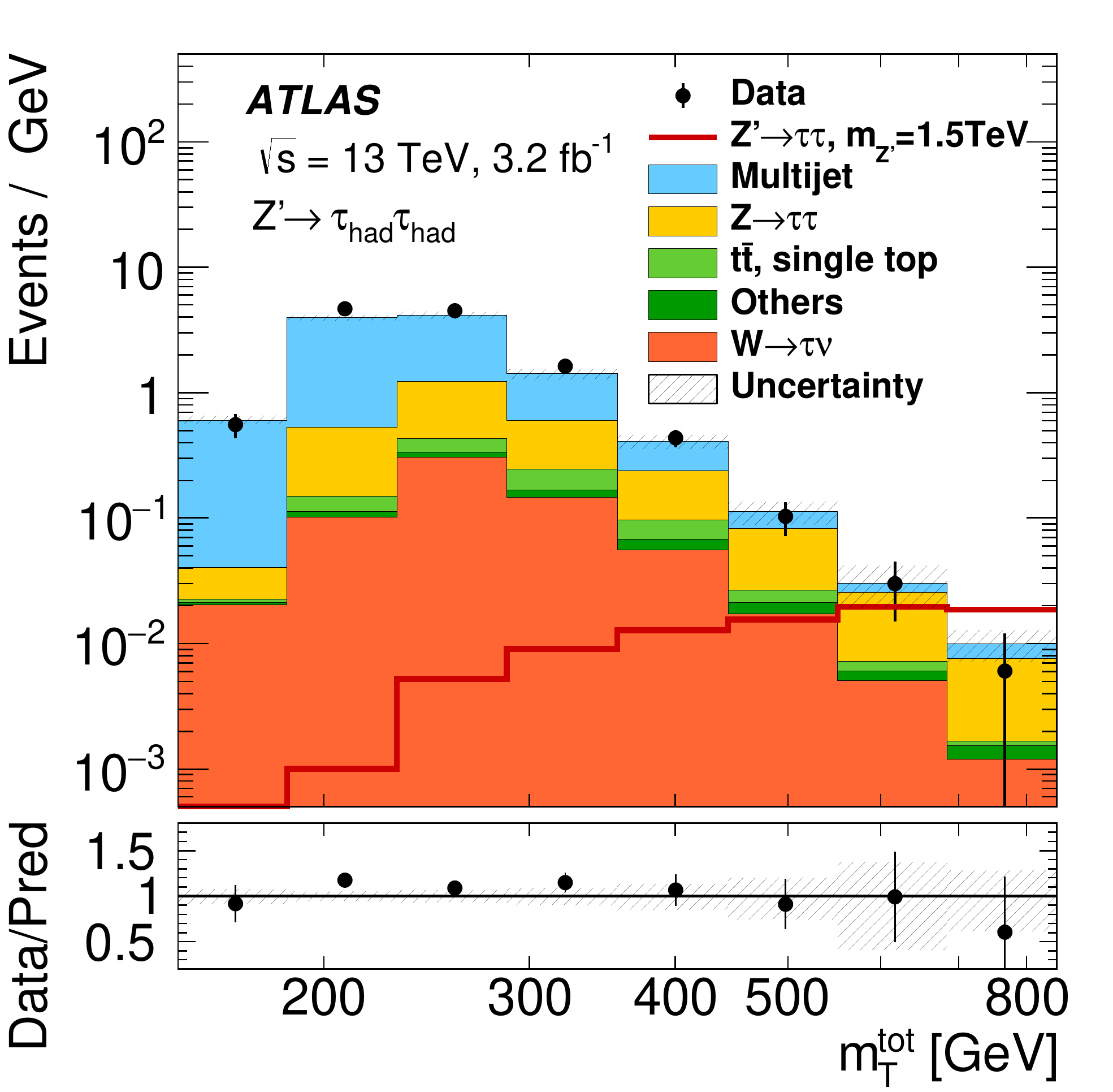} }

  \caption{The distribution of $\mttot$ for the $b$-veto category of the 
    (a) $\taulep\tauhad$  and  (b) $\tauhad\tauhad$  channels, 
    the $b$-tag category of the (c) $\taulep\tauhad$  and (d)
    $\tauhad\tauhad$  channels, and the inclusive category of the 
    (e) $\taulep\tauhad$  and (f) $\tauhad\tauhad$  channels. The label 
    ``Others'' in (b), (d) and (f) refers to contributions
    due to diboson, $Z(\to\ell\ell)$+jets and $W(\to\ell\nu)$+jets production.  
    For the $b$-veto and $b$-tag categories, the binning displayed is that 
    entering into the statistical fit discussed in Section~\ref{sec:results}, 
    while the predictions and uncertainties for the background and signal
    processes are obtained from the fit under the hypothesis of no signal. 
    The inclusive category distributions are shown before any statistical 
    fit.  Overflows are included in the last bin of the distributions.}

\label{fig:mttotPF}
\end{figure}

The fractional contributions of the most important sources of systematic uncertainty to the total uncertainty 
in the signal cross-section measurement are shown for two signal assumptions:
Table~\ref{tab:systematics}(top) represents an MSSM Higgs boson hypothesis
($m_A=500$~\GeV, $\tan\beta=20$) and Table~\ref{tab:systematics}(bottom) corresponds to an SSM \Zprime{}
boson hypothesis ($m_{\Zprime}=1.75$~\TeV). As shown in this table, the sensitivity of the
search is limited by statistical uncertainties.

\begin{table}
  \centering
{  \footnotesize
\begin{tabular}{lrr}
    \multicolumn{3}{c}{MSSM analysis}\\
  \hline
 Source of uncertainty &  $F_{-}$ (\%) & $F_{+}$ (\%)\\
  \hline

  $t\bar{t}$ and single-top-quark backgrounds normalization & $-13\phantom{.0}  $ & $ +11\phantom{.0}$\\
  \tauhadvis{} energy scale                                 & $-\phantom{0}3\phantom{.0}   $ & $ +\phantom{0}8\phantom{.0}$\\
  $\tauhad$ trigger                                         & $-\phantom{0}0.5 $ & $ +10\phantom{.0}$  \\
  Signal acceptance                                         & $-\phantom{0}6\phantom{.0}   $ & $ +\phantom{0}1.9$  \\
  Jet-to-$\tauhadvis$ fake rate  ($\taulep\tauhad$)         & $-\phantom{0}1.5 $ & $ +\phantom{0}2.4$  \\ 
  Multi-jet background ($\tauhad\tauhad$)                   & $-\phantom{0}0.4 $ & $ +\phantom{0}0.3$  \\
  $\ttbar$ background modelling                             & $-\phantom{0}0.1 $ & $ +\phantom{0}1.0$  \\
  Jet-to-$\tauhadvis$ fake rate  ($\tauhad\tauhad$)         & $-\phantom{0}0.2 $ & $ +\phantom{0}0.2$  \\
  Jet-to-vertex association                                 & $-\phantom{0}0.1 $ & $ +\phantom{0}0.1$  \\
  \hline
  Statistics (data and simulation)                          & $-75\phantom{.0}  $ & $ +65\phantom{.0}$ \\
  
  \hline
&  &  \\
&  &  \\

    \multicolumn{3}{c}{\ZprimeSSM{} analysis}\\

    \hline
    Source of uncertainty &  $F_{-}$ (\%) &  $F_{+}$ (\%)\\
    \hline

    \tauhadvis{} energy scale                                    & $-13\phantom{.000}$ & $ +\phantom{0}5\phantom{.00}$ \\
    Pile-up                                                      & $-\phantom{0}0.015 $ & $ +\phantom{0}1.3\phantom{0}$ \\
    $Z+$jets backgrounds cross section and acceptance            & $-\phantom{0}0.3\phantom{00}$ & $ +\phantom{0}0.4\phantom{0}$  \\
    Signal PDF                                                   & $-\phantom{0}0.3\phantom{00}$ & $ +\phantom{0}0.5\phantom{0}$  \\
    Jet-to-$\tauhadvis$ fake rate  ($\tauhad\tauhad$)            & $-\phantom{0}0.3\phantom{00}$ & $ +\phantom{0}0.3\phantom{0}$  \\
    \tauhadvis{} identification                                  & $-\phantom{0}0.19\phantom{0}$ & $ +\phantom{0}0.18$  \\
    Luminosity                                                   & $-\phantom{0}0.16\phantom{0}$ & $ +\phantom{0}0.15$  \\
    $W\to\tau\nu$+jets background reweighting ($\tauhad\tauhad$) & $-\phantom{0}0.17\phantom{0}$ & $ +\phantom{0}0.10$  \\
    \hline
    Statistics  (data and simulation)                            & $-85\phantom{.000}$ & $ +92\phantom{.00}$ \\
    
    \hline

\end{tabular}
}

  \caption{\label{tab:systematics}
Fractional impact of the most important sources of systematic uncertainty on the total 
uncertainty of the signal strength, for
(top) the MSSM signal hypothesis of $m_A=500$~\GeV, $\tan\beta=20$,
(bottom) the \ZprimeSSM{} signal hypothesis of $m_{\Zprime}=1.75$~\TeV. For each source of 
uncertainty,~$F_{\pm} = \pm\frac{ \sigma^{2}_{\textrm{source}} } { \sigma^{2}_{\textrm{total}}}$
is defined as the positive (negative) fractional contribution to the signal strength uncertainty.}
\end{table}

The data are found to be in good agreement with the predicted background yields and hence the
results are given as  exclusion limits.  These are set using the modified frequentist method known
as CL$_{s}$~\cite{CLs_2002}.  Observed and expected 95\% confidence level (CL) upper limits on the
cross section times branching fraction for the production of a single scalar boson $H/A$ decaying to
$\tau\tau$, as a function of the mass of the boson $m_{H/A}$, are shown in
Figures~\ref{fig:limits-crosssection-1}(a) and \ref{fig:limits-crosssection-1}(b).  The limits are
calculated for both the gluon--gluon fusion and $b$-associated production modes, using a combination
of the $\taulep\tauhad$ and $\tauhad\tauhad$ channels and assuming the natural width of the boson to
be negligible compared to the experimental resolution (as expected over the probed MSSM parameter
space).  The lowest excluded cross section times branching fraction values range from $\sigma \times
BR = 1.4~\mathrm{pb}$ at $m_{H/A} = 200~\GeV$ to $\sigma \times BR = 0.025~\mathrm{pb}$ at $m_{H/A}
= 1.2~\TeV$ for a scalar boson produced via gluon--gluon fusion. Similarly, for the $b$-associated
production mechanism the lowest excluded values range is from $\sigma \times BR = 1.6~\mathrm{pb}$ at $m_{H/A} =
200~\GeV$ to $\sigma \times BR = 0.028~\mathrm{pb}$ at $m_{H/A} = 1.2~\TeV$.

\begin{figure}[]
  \centering
  \subfloat[gluon--gluon fusion]{\includegraphics[width=0.45\textwidth]{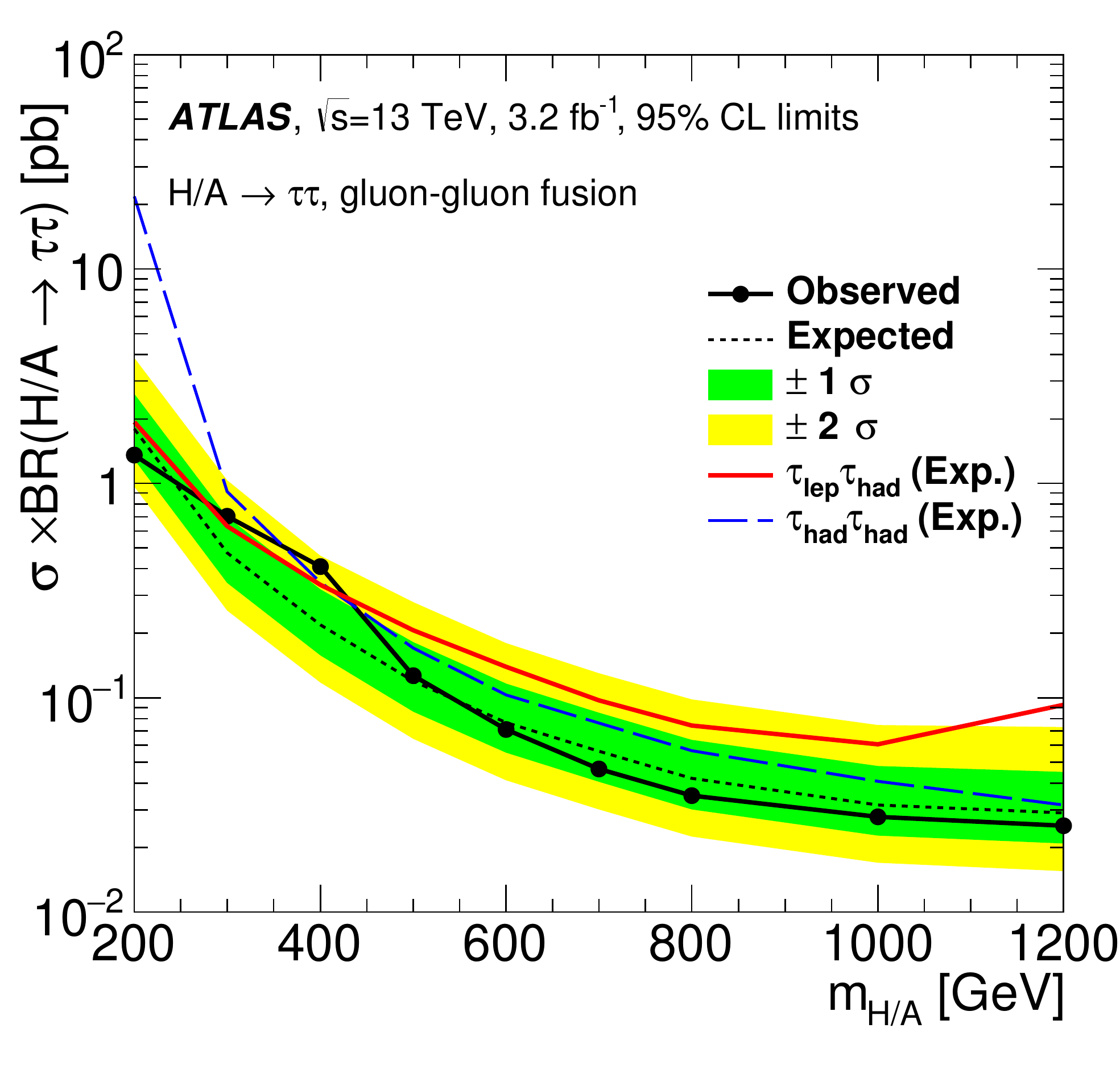} }
  \subfloat[$b$-associated production]{\includegraphics[width=0.45\textwidth]{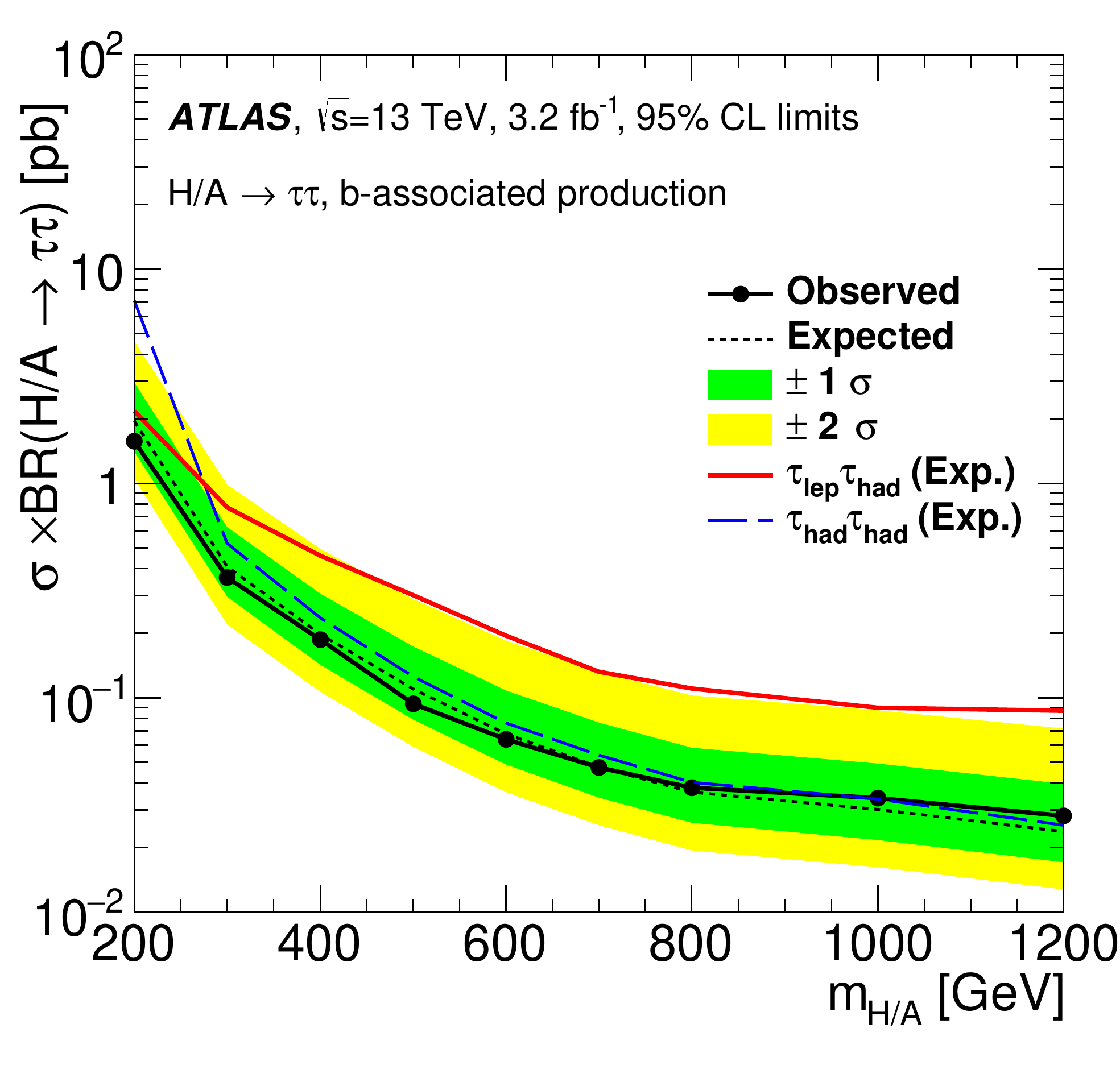} }\\
  \subfloat[$m_h^{\text{mod+}}$ scenario]{\includegraphics[width=0.450\textwidth]{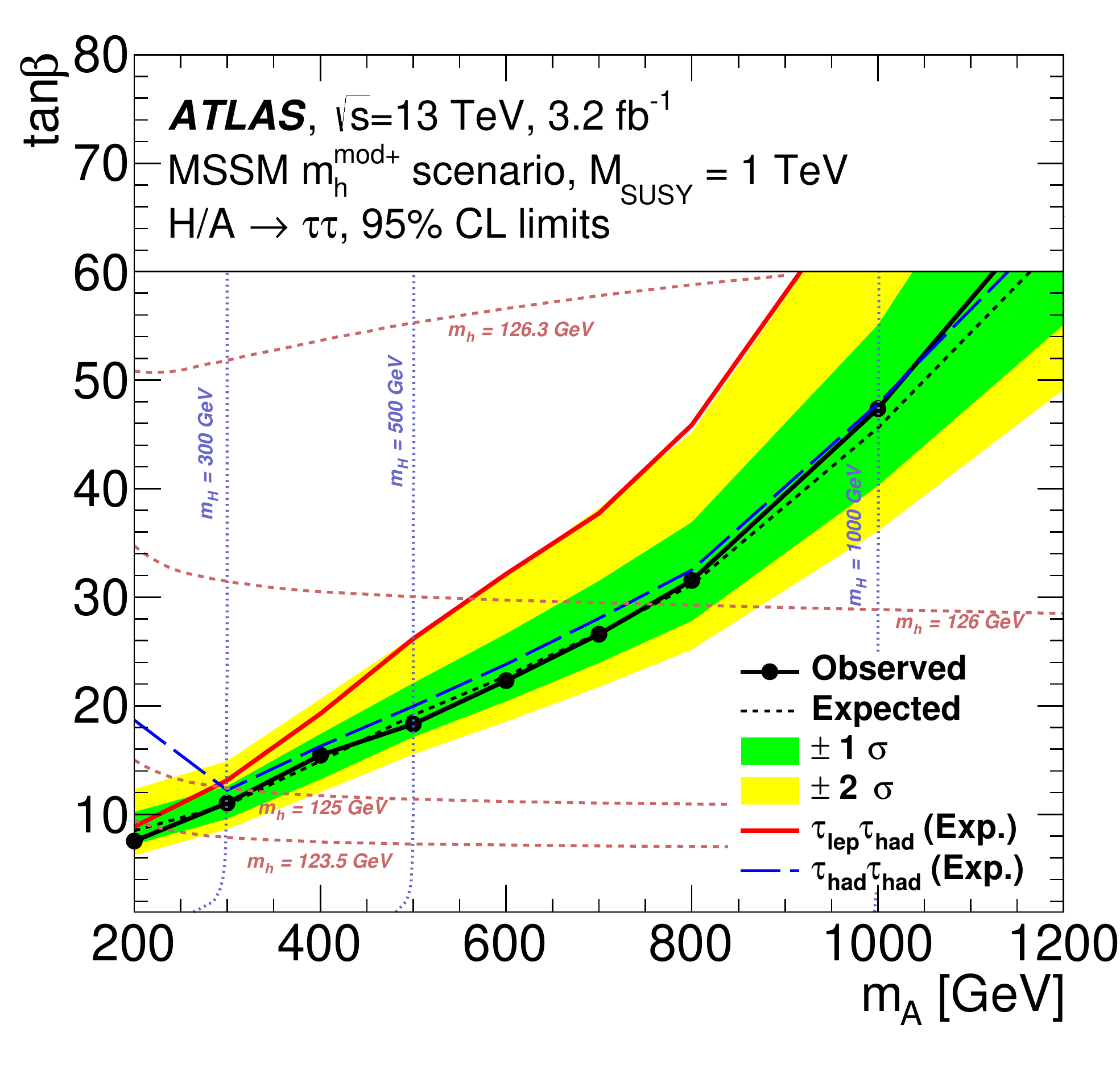}}
  \subfloat[hMSSM scenario]{\includegraphics[width=0.450\textwidth]{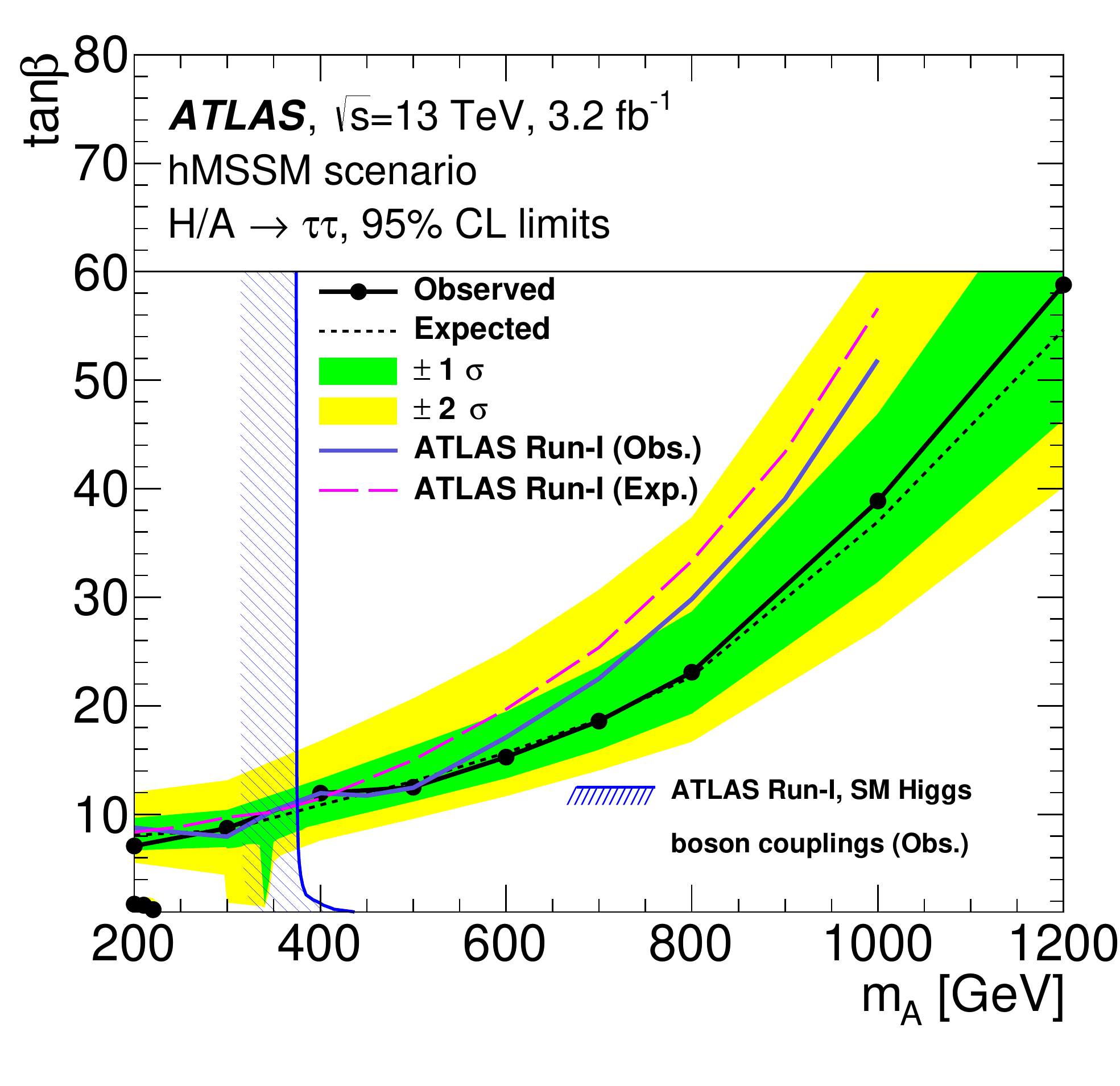}}  \\

  \caption{The observed and expected 95\% CL upper limits on the production 
    cross section times branching fraction of a scalar particle are shown 
    for the combination of the $\taulep\tauhad$ and the $\tauhad\tauhad$ 
    channels.  The production mechanism of $H/A\to\tau\tau$ is assumed to be 
    (a) gluon--gluon fusion or (b) $b$-associated production.
    For comparison, the expected limits for the individual channels, 
    $\taulep\tauhad$ and  $\tauhad\tauhad$, are shown as well.
    The observed and expected 95\% CL  limits on $\tan\beta$ as 
    a function of $m_A$  are shown in (c) for the MSSM $m_h^{\text{mod+}}$ 
    scenario and (d) for the hMSSM scenario.  For comparison, the expected 
    limits from the individual channels, $\taulep\tauhad$ and 
    $\tauhad\tauhad$, are given in (c), while the observed and expected 
    limits from the ATLAS Run-1 analysis in Ref.~\cite{Aad:2014vgg} are
    shown in (d). Dashed lines of constant $m_h$ and $m_H$  are 
    shown in red and blue, respectively.
  }
\label{fig:limits-crosssection-1}
\end{figure}

The observed and expected 95\% CL  limits on
$\tan\beta$ as a function of $m_A$, for the combination of
$\taulep\tauhad$ and $\tauhad\tauhad$ channels in the MSSM
$m_h^{\text{mod+}}$ and hMSSM scenarios are shown in
Figures~\ref{fig:limits-crosssection-1}(c) and \ref{fig:limits-crosssection-1}(d). 
The expected limit in the $m_h^{\text{mod+}}$
scenario is compared to the expected limits from the individual
$\taulep\tauhad$ and $\tauhad\tauhad$ channels. For the
$m_h^{\text{mod+}}$ figure, lines of constant $m_h$ and $m_H$ are shown.
For the hMSSM scenario, the exclusion arising 
from the SM Higgs boson coupling measurements of Ref.~\cite{Aad:2015pla}
is also shown, in addition to the ATLAS Run-1 $H/A\to \tau\tau$ search
result of Ref.~\cite{Aad:2014vgg}.
The $\tan\beta$ constraints in the hMSSM scenario are stronger than those in the $m_h^{\text{mod+}}$
scenario. This is due to the presence of low-mass neutralinos in the $m_h^{\text{mod+}}$ scenario
that reduce the $ H/A\to \tau\tau$ branching fraction and which are absent in the hMSSM scenario. 
In the hMSSM scenario, the
most stringent constraints on $\tan\beta$ for the combined search exclude
$\tan\beta > 7.1$ for $m_A = 200~\GeV$ and $\tan\beta> 39$ for $m_A = 1~\TeV$ at the 95\% CL.
In the MSSM $m_h^{\text{mod+}}$ scenario, the 95\% CL upper limits exclude $\tan\beta > 7.6$ 
for $m_A = 200~\GeV$ and $\tan\beta> 47$ for $m_A = 1~\TeV$.
The feature of the expected limits in the hMSSM scenario exclusion plot at around $m_A = 350$~\GeV{}
is due to the behaviour of the branching ratio $A\to\tau\tau$ close to the $A\to\ttbar$ kinematic 
threshold. Some sensitivity of the search is also expected around $\tan\beta\sim 1$, 
$m_A\sim 200$~\GeV{} due to the increase of the gluon--gluon fusion cross section 
induced by the increased coupling to the top quark.

In the search for the \Zprime{} boson, the observed number of events in the
signal regions of the $\taulep\tauhad$ and $\tauhad\tauhad$ channels
are consistent with the SM predictions. The resulting 95\% CL upper
limits are set on the cross section times branching fraction as a
function of the mass and shown in Figure~\ref{fig:limits}(a). 
These results are interpreted in the context of the
SSM and SFM in Figures~\ref{fig:limits}(a) and \ref{fig:limits}(b), respectively.  The resulting observed (expected)
lower limit on the mass of the \ZPSSM{} boson is $\ZPHHSSMOBSLIMIT{}$ ($\ZPHHSSMEXPLIMIT$)~\TeV.
In the search for the \ZprimeSFM{} boson, results are presented as a function of $\sinsquarephi{}$,
where $\phi$ is the mixing angle between the two SU(2) gauge eigenstates of the model.
Masses below \ZPSFMLOWESTLIMIT{}--\ZPSFMHIGHESTLIMIT{}~\TeV{} are excluded
in the range $0.1 < \sinsquarephi{} < 0.5$, assuming no $\mu-\tau$ mixing. For
the value of $\sinsquarephi{}=0.03$, the lower limit
on the mass of a \ZprimeSFM{} boson is \ZPSFMLIMITLOWSINSQUAREPHI{}~\TeV,
extending the limits from previous direct and indirect searches by more than 200~\GeV.

\begin{figure}[]
  \centering
  \subfloat[\ZprimeSSM{} scenario]{\includegraphics[width=0.450\textwidth]{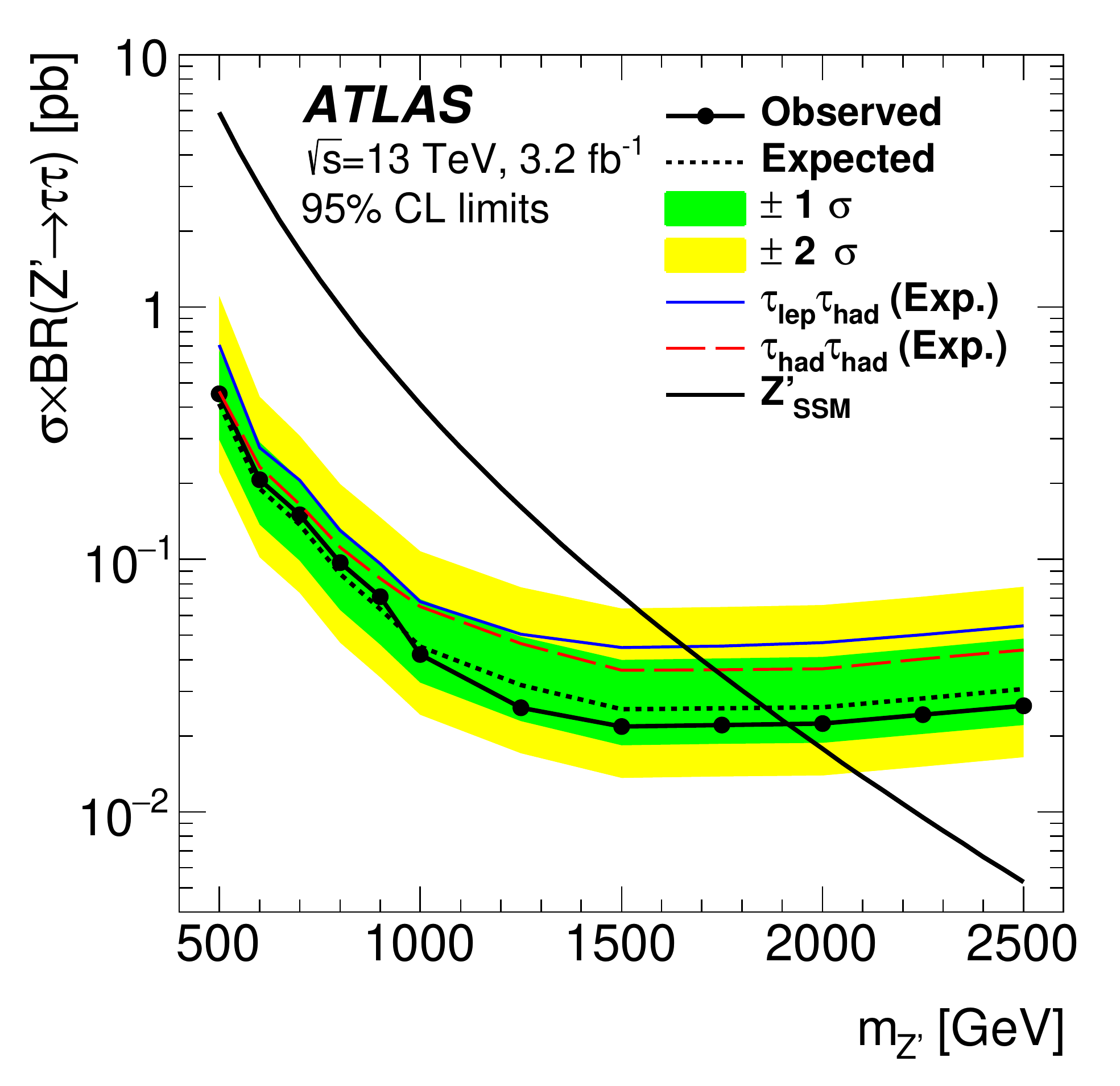}}
  \subfloat[\ZprimeSFM{} scenario]{\includegraphics[width=0.450\textwidth]{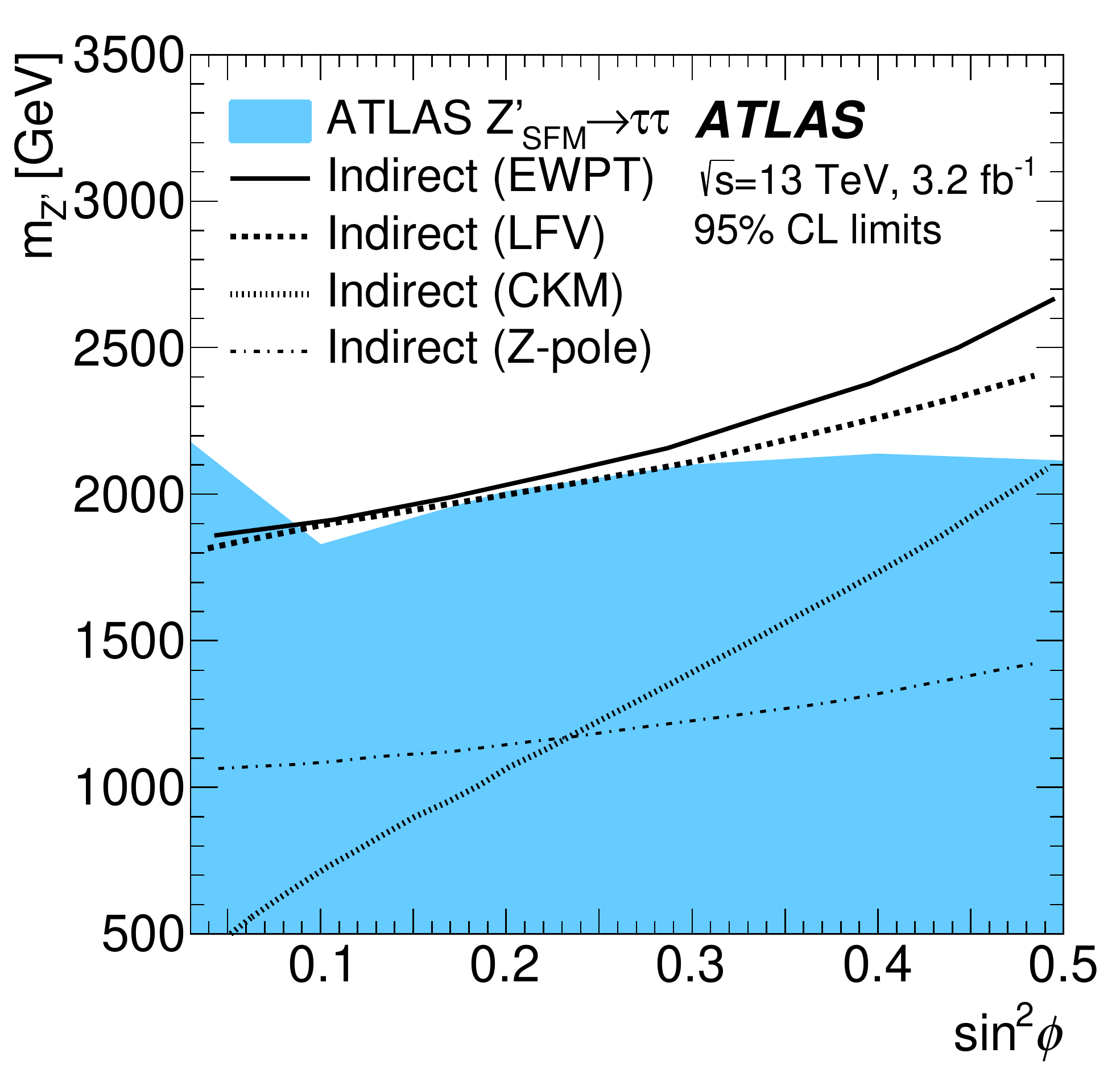}}
  \caption{The 95\% CL upper limit on the cross section times branching fraction for a
    \ZP{} in (a) the Sequential Standard Model and 95\% CL exclusion  
    on (b) the SFM parameter space , overlaid with indirect limits at 95\% CL 
    from fits to electroweak precision measurements~\cite{G221LHC}, lepton 
    flavour violation~\cite{LeeLFV}, CKM
    unitarity~\cite{LeeCKM} and  $Z$-pole measurements~\cite{sfm}.  
  }
\label{fig:limits}
\end{figure}

\clearpage

\section{Conclusions}
\label{sed:conclusions}

A search for neutral Higgs bosons of the Minimal
Supersymmetric Standard Model (MSSM) and for a \Zprime{} gauge boson decaying to a pair of
$\tau$ leptons is performed using a data sample
corresponding to an integrated luminosity of \currentlumirange{} from
 proton--proton collisions at $\sqrt{s} = 13$~\TeV\ recorded by the ATLAS detector at the LHC.
The search finds no indication of an excess over the expected background in the channels
considered. Limits are set at the 95\% CL, which provide constraints in the MSSM
parameter space.  Model-independent upper limits on the production
cross section times the $\tau\tau$ branching fraction of a scalar
boson versus its mass, in both the gluon--gluon fusion and $b$-associated
production modes, are presented. The upper limits on the cross section 
times branching fraction range from
$1.4~(1.6)~\mathrm{pb}$ at $m_{H/A} = 200 \GeV$ to 
$0.025~(0.028)~\mathrm{pb}$ at $m_{H/A} = 1.2 \TeV$ for a scalar boson produced
via gluon--gluon fusion ($b$-associated production). 
In the context of the MSSM $m_{h}^{\text{mod+}}$ scenario, the most stringent 95\%
CL upper limit on $\tan\beta$ for the combined search is $\tan\beta < 7.6$ for $m_A = 200~\GeV$.
This analysis extends the limits of the previous searches for the mass range  $m_A > 500~\GeV$.
The search for a  \Zprime{} boson is interpreted in the context of the 
Sequential Standard Model (SSM) and the Strong Flavour Model (SFM).
Upper limits at the 95\% CL are set on the cross section times branching fraction
as a function of the \Zprime{} mass.
The observed  lower limit on the \Zprime{}  mass is $\ZPHHSSMOBSLIMIT{}$~\TeV{} for a \ZPSSM{}
and ranges from \ZPSFMLOWESTLIMIT{} to \ZPSFMHIGHESTLIMIT{}~\TeV{} as a function of the $\sinsquarephi{}$ parameter
for a \ZprimeSFM{}.

\FloatBarrier
\section*{Acknowledgements}

We thank CERN for the very successful operation of the LHC, as well as the
support staff from our institutions without whom ATLAS could not be
operated efficiently.

We acknowledge the support of ANPCyT, Argentina; YerPhI, Armenia; ARC, Australia; BMWFW and FWF, Austria; ANAS, Azerbaijan; SSTC, Belarus; CNPq and FAPESP, Brazil; NSERC, NRC and CFI, Canada; CERN; CONICYT, Chile; CAS, MOST and NSFC, China; COLCIENCIAS, Colombia; MSMT CR, MPO CR and VSC CR, Czech Republic; DNRF and DNSRC, Denmark; IN2P3-CNRS, CEA-DSM/IRFU, France; GNSF, Georgia; BMBF, HGF, and MPG, Germany; GSRT, Greece; RGC, Hong Kong SAR, China; ISF, I-CORE and Benoziyo Center, Israel; INFN, Italy; MEXT and JSPS, Japan; CNRST, Morocco; FOM and NWO, Netherlands; RCN, Norway; MNiSW and NCN, Poland; FCT, Portugal; MNE/IFA, Romania; MES of Russia and NRC KI, Russian Federation; JINR; MESTD, Serbia; MSSR, Slovakia; ARRS and MIZ\v{S}, Slovenia; DST/NRF, South Africa; MINECO, Spain; SRC and Wallenberg Foundation, Sweden; SERI, SNSF and Cantons of Bern and Geneva, Switzerland; MOST, Taiwan; TAEK, Turkey; STFC, United Kingdom; DOE and NSF, United States of America. In addition, individual groups and members have received support from BCKDF, the Canada Council, CANARIE, CRC, Compute Canada, FQRNT, and the Ontario Innovation Trust, Canada; EPLANET, ERC, FP7, Horizon 2020 and Marie Sk{\l}odowska-Curie Actions, European Union; Investissements d'Avenir Labex and Idex, ANR, R{\'e}gion Auvergne and Fondation Partager le Savoir, France; DFG and AvH Foundation, Germany; Herakleitos, Thales and Aristeia programmes co-financed by EU-ESF and the Greek NSRF; BSF, GIF and Minerva, Israel; BRF, Norway; Generalitat de Catalunya, Generalitat Valenciana, Spain; the Royal Society and Leverhulme Trust, United Kingdom.

The crucial computing support from all WLCG partners is acknowledged gratefully, in particular from CERN, the ATLAS Tier-1 facilities at TRIUMF (Canada), NDGF (Denmark, Norway, Sweden), CC-IN2P3 (France), KIT/GridKA (Germany), INFN-CNAF (Italy), NL-T1 (Netherlands), PIC (Spain), ASGC (Taiwan), RAL (UK) and BNL (USA), the Tier-2 facilities worldwide and large non-WLCG resource providers. Major contributors of computing resources are listed in Ref.~\cite{ATL-GEN-PUB-2016-002}.

\printbibliography
\clearpage
\begin{flushleft}
{\Large The ATLAS Collaboration}

\bigskip

M.~Aaboud$^\textrm{\scriptsize 135d}$,
G.~Aad$^\textrm{\scriptsize 86}$,
B.~Abbott$^\textrm{\scriptsize 113}$,
J.~Abdallah$^\textrm{\scriptsize 8}$,
O.~Abdinov$^\textrm{\scriptsize 12}$,
B.~Abeloos$^\textrm{\scriptsize 117}$,
R.~Aben$^\textrm{\scriptsize 107}$,
O.S.~AbouZeid$^\textrm{\scriptsize 137}$,
N.L.~Abraham$^\textrm{\scriptsize 149}$,
H.~Abramowicz$^\textrm{\scriptsize 153}$
% ...
\bigskip
\\
% ...
$^{8}$ Department of Physics, The University of Texas at Arlington, Arlington TX, United States of America\\
% ...
$^{12}$ Institute of Physics, Azerbaijan Academy of Sciences, Baku, Azerbaijan\\
% ...
$^{86}$ CPPM, Aix-Marseille Universit{\'e} and CNRS/IN2P3, Marseille, France\\
% ...
$^{107}$ Nikhef National Institute for Subatomic Physics and University of Amsterdam, Amsterdam, Netherlands\\
% ...
$^{113}$ Homer L. Dodge Department of Physics and Astronomy, University of Oklahoma, Norman OK, United States of America\\
% ...
$^{117}$ LAL, Univ. Paris-Sud, CNRS/IN2P3, Universit{\'e} Paris-Saclay, Orsay, France\\
% ...
$^{135}$ $^{(a)}$ Facult{\'e} des Sciences Ain Chock, R{\'e}seau Universitaire de Physique des Hautes Energies - Universit{\'e} Hassan II, Casablanca; $^{(b)}$ Centre National de l'Energie des Sciences Techniques Nucleaires, Rabat; $^{(c)}$ Facult{\'e} des Sciences Semlalia, Universit{\'e} Cadi Ayyad, LPHEA-Marrakech; $^{(d)}$ Facult{\'e} des Sciences, Universit{\'e} Mohamed Premier and LPTPM, Oujda; $^{(e)}$ Facult{\'e} des sciences, Universit{\'e} Mohammed V, Rabat, Morocco\\
% ...
$^{137}$ Santa Cruz Institute for Particle Physics, University of California Santa Cruz, Santa Cruz CA, United States of America\\
% ...
$^{149}$ Department of Physics and Astronomy, University of Sussex, Brighton, United Kingdom\\
% ...
$^{153}$ Raymond and Beverly Sackler School of Physics and Astronomy, Tel Aviv University, Tel Aviv, Israel\\
% ...
$^{a}$ Also at Department of Physics, King's College London, London, United Kingdom\\
$^{b}$ Also at Institute of Physics, Azerbaijan Academy of Sciences, Baku, Azerbaijan\\
$^{c}$ Also at Novosibirsk State University, Novosibirsk, Russia\\
$^{d}$ Also at TRIUMF, Vancouver BC, Canada\\
$^{e}$ Also at Department of Physics {\&} Astronomy, University of Louisville, Louisville, KY, United States of America\\
% ...
\end{flushleft}

\end{document}